\documentclass{ws-ijgmmp}

\begin{document}

\def\II{{\rm 1\!\hskip-1pt I}}
\def\cstok#1{\leavevmode\thinspace\hbox{\vrule\vtop{\vbox
{\hrule\kern1pt
\hbox{\vphantom{\tt/}\thinspace{\tt#1}\thinspace}}
\kern1pt\hrule}\vrule}\thinspace}

\markboth{Giampiero Esposito}
{From Spinor Geometry to Complex General Relativity}

\catchline{}{}{}{}{}

\title{FROM SPINOR GEOMETRY TO COMPLEX GENERAL
RELATIVITY\footnote{With kind permission of Springer Science and
Business Media to use the material described in the Acknowledgments.}}

\author{GIAMPIERO ESPOSITO}

\address{INFN, Sezione di Napoli, and Dipartimento di Scienze
Fisiche, Universit\`a Federico II, Complesso Universitario di
Monte S. Angelo, Via Cintia, Edificio N'\\
80126 Napoli, Italy\, \\
\email{giampiero.esposito@na.infn.it}}

\maketitle

\begin{history}
\received{(1 April 2005)}
\end{history}

\begin{abstract}
An attempt is made of giving a self-contained
introduction to holomorphic ideas in general relativity, following
work over the last thirty years by several authors. The main topics
are complex manifolds, two-component spinor calculus, conformal
gravity, $\alpha$-planes in Minkowski space-time, 
$\alpha$-surfaces and twistor geometry, anti-self-dual space-times
and Penrose transform, spin-3/2 potentials, heaven spaces and heavenly
equations.
\end{abstract}


\section{Introduction to Complex Space-Time}

The physical
and mathematical motivations for studying complex space-times
or real Riemannian four-manifolds in gravitational physics
are first described.
They originate from algebraic geometry, Euclidean quantum
field theory, the path-integral approach to quantum
gravity, and the theory of conformal gravity.
The theory of complex manifolds is then briefly outlined.
Here, one deals with paracompact Hausdorff spaces where
local coordinates transform by complex-analytic
transformations. Examples are given such as complex
projective space $P_{m}$, non-singular sub-manifolds
of $P_{m}$, and orientable surfaces. The
plan of the whole paper is eventually presented, with
emphasis on two-component spinor calculus, Penrose transform
and Penrose formalism for spin-${3\over 2}$ potentials.

\subsection{From Lorentzian to complex space-time}

Although Lorentzian geometry is the mathematical framework
of classical general relativity and can be seen as a good
model of the world we live in (\cite{HaEl73}, \cite{Espo92},
\cite{Espo94}), the theoretical-physics
community has developed instead many models based on a complex
space-time picture. We postpone until Sec. 3.3 the
discussion of real, complexified or complex manifolds, and
we here limit ourselves to say that the main motivations for studying
these ideas are as follows.

(1) When one tries to make sense of quantum field theory
in flat space-time, one finds it very convenient to
study the Wick-rotated version of Green functions,
since this leads to well defined mathematical calculations
and elliptic boundary-value problems. At the end, quantities
of physical interest are evaluated by analytic continuation
back to {\it real} time in Minkowski space-time.

(2) The singularity at $r=0$ of the Lorentzian Schwarzschild 
solution disappears on the real Riemannian section of the
corresponding complexified space-time, since $r=0$ no longer
belongs to this manifold (\cite{Espo94}). Hence there are
real Riemannian four-manifolds which are singularity-free,
and it remains to be seen whether they are the most 
fundamental in modern theoretical physics.

(3) Gravitational instantons shed some light on possible
boundary conditions relevant for path-integral quantum
gravity and quantum cosmology (\cite{Hawk79}, \cite{GiHa93},
\cite{Espo94}).

(4) Unprimed and primed spin-spaces 
are not (anti-)isomorphic
if Lorentzian space-time is replaced by a complex or real
Riemannian manifold. Thus, for example, the Maxwell field
strength is represented by two independent symmetric spinor
fields, and the Weyl curvature is also represented by two
independent symmetric spinor fields (see (2.1.35) and (2.1.36)). 
Since such spinor
fields are no longer related by complex conjugation
(i.e. the (anti-)isomorphism between the two spin-spaces), one of
them may vanish without the other one having to vanish
as well. This property gives rise to the so-called self-dual
or anti-self-dual gauge fields, as well as to self-dual or
anti-self-dual space-times (Sec. 4.2).

(5) The geometric study of this special class of space-time
models has made substantial progress by using twistor-theory
techniques. The underlying idea (\cite{Penr67}, 
\cite{Penr68}, \cite{PeMa73}, \cite{Penr75}, \cite{Penr77},
\cite{Penr80}, \cite{PeWa80}, 
\cite{Ward80a}, \cite{Ward80b}, \cite{Penr81}, \cite{Ward81a},
\cite{Ward81b}, \cite{Hugg85}, \cite{HuTo85}, \cite{Wood85},
\cite{Penr86}, \cite{Penr87}, \cite{Yass87}, \cite{Mani88},
\cite{BaBa90}, \cite{MaHu90}, \cite{WaWe90}, \cite{MaWo96})
is that conformally invariant concepts
such as null lines and null surfaces are the basic
building blocks of the world we live in, whereas space-time
points should only appear as a derived concept. By using
complex-manifold theory, twistor theory provides an
appropriate mathematical description of this key idea.

A possible mathematical motivation for twistors can be
described as follows (papers 99 and 100 in \cite{Atiy88}). In
two real dimensions, many interesting problems are best tackled
by using complex-variable methods. In four real dimensions,
however, the introduction of two complex coordinates is not,
by itself, sufficient, since no preferred choice exists. In other
words, if we define the complex variables
$$
z_{1} \equiv x_{1}+ix_{2},
\eqno (1.1.1)
$$
$$
z_{2} \equiv x_{3}+ix_{4},
\eqno (1.1.2)
$$
we rely too much on this particular coordinate system, and a
permutation of the four real coordinates $x_{1},x_{2},x_{3},x_{4}$
would lead to new complex variables not well related to the
first choice. One is thus led to introduce three complex variables
$\Bigr(u,z_{1}^{u},z_{2}^{u}\Bigr)$: the first variable $u$ tells
us which complex structure to use, and the next two are the
complex coordinates themselves. In geometric language, we start
with the complex projective three-space $P_{3}(C)$ (see Sec. 1.2)
with complex homogeneous coordinates $(x,y,u,v)$, and we remove the
complex projective line given by $u=v=0$. Any line in
$\Bigr(P_{3}(C)-P_{1}(C)\Bigr)$ is thus given by a pair of
equations
$$
x=au+bv,
\eqno (1.1.3)
$$
$$
y=cu+dv.
\eqno (1.1.4)
$$
In particular, we are interested in those lines for which
$c=-{\overline b},d={\overline a}$. The determinant $\Delta$ of
(1.1.3) and (1.1.4) is thus given by
$$
\Delta=a{\overline a}+b{\overline b}={|a|}^{2}
+{|b|}^{2},
\eqno (1.1.5)
$$
which implies that the line given above never intersects
the line $x=y=0$, with the obvious exception of the case when 
they coincide. Moreover, no two lines intersect, and they fill
out the whole of $\Bigr(P_{3}(C)-P_{1}(C)\Bigr)$. This leads to
the fibration $\Bigr(P_{3}(C)-P_{1}(C)\Bigr)\longrightarrow
R^{4}$ by assigning to each point of $\Bigr(P_{3}(C)-P_{1}(C)\Bigr)$
the four coordinates $\Bigr({\rm Re}(a),{\rm Im}(a),
{\rm Re}(b),{\rm Im}(b)\Bigr)$.
Restriction of this fibration to a plane of the form
$$
\alpha u + \beta v =0,
\eqno (1.1.6)
$$
yields an isomorphism $C^{2} \cong R^{4}$, which depends on the
ratio $(\alpha,\beta) \in P_{1}(C)$. This is why the picture
embodies the idea of introducing complex coordinates.

Such a fibration depends on the conformal structure of $R^{4}$.
Hence, it can be extended to the one-point compactification
$S^{4}$ of $R^{4}$, so that we get a fibration $P_{3}(C)
\longrightarrow S^{4}$ where the line $u=v=0$, previously 
excluded, sits over the point at $\infty$ of
$S^{4}=R^{4} \cup \Bigr \{ \infty \Bigr \}$. This fibration is
naturally obtained if we use the quaternions $H$ to identify
$C^{4}$ with $H^{2}$ and the four-sphere $S^{4}$ with $P_{1}(H)$,
the quaternion projective line. We should now recall that the
quaternions $H$ are obtained from the vector space $R$ of real
numbers by adjoining three symbols $i,j,k$ such that
$$
i^{2}=j^{2}=k^{2}=-1,
\eqno (1.1.7)
$$
$$
ij=-ji=k, \; 
jk=-kj=i, \; 
ki=-ik=j .
\eqno (1.1.8)
$$
Thus, a general {\it quaternion} $\in H$ is defined by
$$
x \equiv x_{1}+x_{2}i+x_{3}j+x_{4}k,
\eqno (1.1.9)
$$
where $\Bigr(x_{1},x_{2},x_{3},x_{4}\Bigr) \in R^{4}$,
whereas the conjugate quaternion $\overline x$ is given by
$$
{\overline x} \equiv x_{1}-x_{2}i-x_{3}j-x_{4}k .
\eqno (1.1.10)
$$
Note that conjugation obeys the identities
$$
{\overline {(xy)}}={\overline y} \; {\overline x},
\eqno (1.1.11)
$$
$$
x {\overline x}={\overline x} x =\sum_{\mu=1}^{4}
x_{\mu}^{2} \equiv {|x|}^{2}.
\eqno (1.1.12)
$$
If a quaternion does not vanish, it has a unique inverse
given by
$$
x^{-1} \equiv {{\overline x} \over {|x|}^{2}}.
\eqno (1.1.13)
$$
Interestingly, if we identify $i$ with $\sqrt{-1}$,
we may view the complex numbers $C$ as
contained in $H$ taking $x_{3}=x_{4}=0$. Moreover,
every quaternion $x$ as in (1.1.9) has a unique
decomposition
$$
x=z_{1}+z_{2}j,
\eqno (1.1.14)
$$
where $z_{1} \equiv x_{1}+x_{2}i$, $z_{2} \equiv x_{3}+x_{4}i$,
by virtue of (1.1.8). This property enables one to identify
$H$ with $C^{2}$, and finally $H^{2}$ with $C^{4}$, as we
said following (1.1.6).

The map $\sigma: P_{3}(C) \longrightarrow P_{3}(C)$
defined by
$$
\sigma(x,y,u,v)=(-{\overline y},{\overline x},-{\overline v},
{\overline u}),
\eqno (1.1.15)
$$
preserves the fibration because $c=-{\overline b},
d={\overline a}$, and induces the antipodal map on each
fibre. We can now lift problems from $S^{4}$ or $R^{4}$
to $P_{3}(C)$ and try to use complex methods.

\subsection{Complex manifolds}

Following \cite{Cher79}, we now describe some basic ideas and
properties of complex-manifold theory. The reader should
thus find it easier (or, at least, less difficult) to
understand the holomorphic ideas used in the rest of the
paper.

We know that a manifold is a space which is locally similar
to Euclidean space in that it can be covered by coordinate
patches. More precisely (\cite{HaEl73}), we say
that a {\it real} $C^{r}$ $n$-dimensional manifold 
$\cal M$ is a set $\cal M$ together with a $C^{r}$ atlas
$\Bigr \{U_{\alpha},\phi_{\alpha} \Bigr \}$, i.e. a collection
of charts $\Bigr(U_{\alpha},\phi_{\alpha}\Bigr)$, where the
$U_{\alpha}$ are subsets of $\cal M$ and the $\phi_{\alpha}$
are one-to-one maps of the corresponding $U_{\alpha}$ into open
sets in $R^{n}$ such that

(i) $\cal M$ is covered by the $U_{\alpha}$, i.e.
${\cal M}=\bigcup_{\alpha} U_{\alpha}$

(ii) if $U_{\alpha} \cap U_{\beta}$ is non-empty, the map
$$
\phi_{\alpha} \circ \phi_{\beta}^{-1}:
\phi_{\beta} \Bigr(U_{\alpha} \cap U_{\beta}\Bigr)
\rightarrow \phi_{\alpha}\Bigr(U_{\alpha}\cap U_{\beta}
\Bigr)
$$
is a $C^{r}$ map of an open subset of $R^{n}$ into an open
subset of $R^{n}$. In general relativity, it is of considerable
importance to require that the Hausdorff separation axiom
should hold. This states that if $p,q$ are any two distinct
points in $\cal M$, there exist disjoint open sets $U,V$ in
$\cal M$ such that $p \in U$, $q \in V$. The space-time
manifold $(M,g)$ is therefore taken to be a connected,
four-dimensional, Hausdorff $C^{\infty}$ manifold $M$ with a
Lorentz metric $g$ on $M$, i.e. the assignment of a 
symmetric, non-degenerate bilinear form 
$g_{\mid p}:T_{p}M \times T_{p}M \rightarrow R$ with diagonal
form $(-,+,+,+)$ to each tangent space. Moreover, a time
orientation is given by a globally defined, timelike vector
field $X:M\rightarrow TM$. This enables one to say that a
timelike or null tangent vector $v \in T_{p}M$ is
future-directed if $g(X(p),v)<0$, or past-directed if
$g(X(p),v)>0$ (\cite{Espo92}, \cite{Espo94}).

By a complex manifold we mean a paracompact Hausdorff space
covered by neighbourhoods each homeomorphic to an open set
in $C^{m}$, such that where two neighbourhoods overlap, the
local coordinates transform by a complex-analytic 
transformation. Thus, if $z^{1},...,z^{m}$ are local coordinates
in one such neighbourhood, and if $w^{1},...,w^{m}$ are local
coordinates in another neighbourhood, where they are both 
defined one has $w^{i}=w^{i}\Bigr(z^{1},...,z^{m}\Bigr)$, where
each $w^{i}$ is a holomorphic function of the $z$'s, and the
determinant $\partial \Bigr(w^{1},...,w^{m}\Bigr)/
\partial \Bigr(z^{1},...,z^{m}\Bigr)$ does not vanish. Various
examples can be given as follows \cite{Cher79}.
\vskip 0.3cm
\noindent
{\bf E1.} The space $C^{m}$ whose points are the 
$m$-tuples of complex numbers $\Bigr(z^{1},...,z^{m}\Bigr)$.
In particular, $C^{1}$ is the so-called Gaussian plane.
\vskip 0.3cm
\noindent
{\bf E2.} Complex projective space $P_{m}$, also denoted by
$P_{m}(C)$ or $CP^{m}$. Denoting by $\{ 0 \}$ the origin
$(0,...,0)$, this is the quotient space obtained by identifying
the points $\Bigr(z^{0},z^{1},...,z^{m}\Bigr)$ in 
$C^{m+1}- \{ 0 \}$ which differ from each other by a factor. The
covering of $P_{m}$ is given by $m+1$ open sets $U_{i}$ defined
respectively by $z^{i} \not =0$, $0 \leq i \leq m$. In $U_{i}$
we have the local coordinates $\zeta_{i}^{k} \equiv
z^{k}/z^{i}$, $0 \leq k \leq m$, $k \not = i$.
In $U_{i} \cap U_{j}$, transition of local coordinates is
given by $\zeta_{j}^{h} \equiv \zeta_{i}^{h}/ \zeta_{i}^{j}$,
$0 \leq h \leq m$, $h \not =j$, which are holomorphic
functions. A particular case is the Riemann sphere $P_{1}$.
\vskip 0.3cm
\noindent
{\bf E3.} Non-singular sub-manifolds of $P_{m}$, in particular,
the non-singular hyperquadric
$$
{\Bigr(z^{0}\Bigr)}^{2}+...+{\Bigr(z^{m}\Bigr)}^{2}=0.
\eqno (1.2.1)
$$
A theorem of Chow states that every compact sub-manifold
embedded in $P_{m}$ is the locus defined by a finite 
number of homogeneous polynomial equations. Compact 
sub-manifolds of $C^{m}$ are not very important, since
a connected compact sub-manifold of $C^{m}$ is a point.
\vskip 0.3cm
\noindent
{\bf E4.} Let $\Gamma$ be the discontinuous group generated
by $2m$ translations of $C^{m}$, which are linearly 
independent over the reals. The quotient space 
$C^{m}/ \Gamma$ is then called the complex torus. Moreover,
let $\Delta$ be the discontinuous group generated by
$z^{k} \rightarrow 2z^{k}$, $1 \leq k \leq m$. The quotient
manifold ${\Bigr(C^{m}- \{0 \} \Bigr)}/ \Delta$ is the
so-called Hopf manifold, and is homeomorphic to
$S^{1} \times S^{2m-1}$. Last but not least, we consider the
group $M_{3}$ of all matrices
$$
E_{3}= \pmatrix {1&z_{1}&z_{2}\cr
0&1&z_{3}\cr 0&0&1\cr},
\eqno (1.2.2)
$$
and let $D$ be the discrete group consisting of those
matrices for which $z_{1},z_{2},z_{3}$ are Gaussian
integers. This means that $z_{k}=m_{k}+in_{k}$,
$1 \leq k \leq 3$, where $m_{k},n_{k}$ are rational
integers. An Iwasawa manifold is then defined as the
quotient space $M_{3}/D$.
\vskip 0.3cm
\noindent
{\bf E5.} Orientable surfaces are particular complex
manifolds. The surfaces are taken to be $C^{\infty}$,
and we define on them a positive-definite Riemannian
metric. The Korn--Lichtenstein theorem ensures that local
parameters $x,y$ exist such that the metric locally takes
the form
$$
g=\lambda^{2} \Bigr(dx \otimes dx + dy \otimes dy \Bigr),
\; \lambda>0,
\eqno (1.2.3)
$$
or
$$
g=\lambda^{2} dz \otimes d{\overline z},
\; z \equiv x+iy.
\eqno (1.2.4)
$$
If $w$ is another local coordinate, we have
$$
g=\lambda^{2} dz \otimes d{\overline z}
=\mu^{2} dw \otimes d{\overline w},
\eqno (1.2.5)
$$
since $g$ is globally defined. Hence $dw$ is a multiple of
$dz$ or $d{\overline z}$. In particular, if the complex
coordinates $z$ and $w$ define the same orientation, then
$dw$ is proportional to $dz$. Thus, $w$ is a holomorphic
function of $z$, and the surface becomes a complex manifold.
Riemann surfaces are, by definition, one-dimensional complex
manifolds.

Let us denote by $V$ an $m$-dimensional real vector space.
We say that $V$ has a {\it complex structure} 
if there exists a linear
endomorphism $J:V \rightarrow V$ such that $J^{2}=-\II$, where
$\II$ is the identity endomorphism. An eigenvalue of $J$ is a 
complex number $\lambda$ such that the equation $Jx=\lambda x$
has a non-vanishing solution $x \in V$. Applying $J$ to both
sides of this equation, one finds $-x=\lambda^{2}x$. Hence
$\lambda=\pm i$. Since the complex eigenvalues occur in
conjugate pairs, $V$ is of even dimension $n=2m$. Let us now
denote by $V^{*}$ the dual space of $V$, i.e. the space of all
real-valued linear functions over $V$. The pairing of $V$ and
$V^{*}$ is $\langle x,y^{*}\rangle$, 
$x \in V$, $y^{*} \in V^{*}$, so that
this function is $R$-linear in each of the arguments. Following
Chern 1979, we also consider $V^{*} \otimes C$, i.e. the space
of all complex-valued $R$-linear functions over $V$. By
construction, $V^{*} \otimes C$ is an $n$-complex-dimensional
complex vector space. Elements $f \in V^{*} \otimes C$ are
of type $(1,0)$ if $f(Jx)=if(x)$, and of type $(0,1)$
if $f(Jx)=-if(x)$, $x \in V$.

If $V$ has a complex structure $J$, an {\it Hermitian structure}
in $V$ is a complex-valued function $H$ acting on $x,y \in V$
such that
$$
H \Bigr(\lambda_{1}x_{1}+\lambda_{2}x_{2},y\Bigr)
=\lambda_{1} H(x_{1},y)+\lambda_{2}
H(x_{2},y)
\; \; x_{1},x_{2},y \in V
\; \; 
\lambda_{1},\lambda_{2} \in R ,
\eqno (1.2.6)
$$
$$
{\overline {H(x,y)}}=H(y,x),
\eqno (1.2.7)
$$
$$
H(Jx,y)=iH(x,y) \Longleftrightarrow
H(x,Jy)=-iH(x,y).
\eqno (1.2.8)
$$
By using the split of $H(x,y)$ into its real and imaginary
parts
$$
H(x,y)=F(x,y)+iG(x,y),
\eqno (1.2.9)
$$
conditions (1.2.7) and (1.2.8) may be re-expressed as
$$
F(x,y)=F(y,x), \;
G(x,y)=-G(y,x),
\eqno (1.2.10)
$$
$$
F(x,y)=G(Jx,y), \;
G(x,y)=-F(Jx,y).
\eqno (1.2.11)
$$

If $\cal M$ is a $C^{\infty}$ manifold of dimension $n$,
and if $T_{x}$ and $T_{x}^{*}$ are tangent and cotangent
spaces respectively at $x \in {\cal M}$, an 
{\it almost complex structure} on $\cal M$ is a
$C^{\infty}$ field of endomorphisms $J_{x}:T_{x}
\rightarrow T_{x}$ such that $J_{x}^{2}=-\II_{x}$, 
where $\II_{x}$ is the identity endomorphism in $T_{x}$.
A manifold with an almost complex structure is called
{\it almost complex}. If a manifold is almost complex,
it is even-dimensional and orientable. However, this
is only a necessary condition. Examples can be found
(e.g. the four-sphere $S^{4}$) of even-dimensional,
orientable manifolds which cannot be given an almost 
complex structure.

\subsection{An outline of this work}

Since this paper is devoted to the geometry of complex
space-time in spinor form, Sec. 2 presents the
basic ideas, methods and results of two-component 
spinor calculus. Such a calculus is described in terms
of spin-space formalism, i.e. a complex vector space
endowed with a symplectic form and some fundamental 
isomorphisms. These mathematical
properties enable one to raise and lower indices, define
the conjugation of spinor fields in Lorentzian or
Riemannian four-geometries, translate tensor fields into
spinor fields (or the other way around). The standard
two-spinor form of the Riemann curvature tensor is then obtained
by relying on the (more) familiar tensor properties of the
curvature. The introductory analysis ends with the Petrov
classification of space-times, expressed in terms of the
Weyl spinor of conformal gravity.

Since the whole of twistor theory may be viewed as a holomorphic 
description of space-time geometry in a conformally invariant
framework, Sec. 3 studies the key results of conformal
gravity, i.e. $C$-spaces, Einstein spaces and complex Einstein
spaces. Hence a necessary and sufficient condition for a space-time
to be conformal to a complex Einstein space is obtained,
following \cite{Koza85}. Such a condition involves the
Bach and Eastwood--Dighton spinors, and their occurrence is
derived in detail. The difference between Lorentzian space-times,
Riemannian four-spaces, complexified space-times and complex
space-times is also analyzed.

Section 4 is a pedagogical introduction to twistor spaces,
from the point of view of mathematical physics and relativity
theory. This is obtained by defining twistors as $\alpha$-planes
in complexified compactified Minkowski space-time, and as 
$\alpha$-surfaces in curved space-time. In the former case, one
deals with totally null two-surfaces, in that the complexified
Minkowski metric vanishes on any pair of null tangent vectors
to the surface. Hence such null tangent vectors have the
form $\lambda^{A}\pi^{A'}$, where $\lambda^{A}$ is varying
and $\pi^{A'}$ is covariantly constant. This definition can be
generalized to complex or real Riemannian four-manifolds, provided
that the Weyl curvature is anti-self-dual. An alternative definition of
twistors in Minkowski space-time is instead based on the vector
space of solutions of a differential equation, which involves
the symmetrized covariant derivative of an unprimed spinor field.
Interestingly, a deep correspondence exists between flat space-time
and twistor space. Hence complex space-time points correspond to
spheres in the so-called projective twistor space, and this
concept is carefully formulated. Sheaf cohomology is then presented
as the mathematical tool necessary to describe a conformally
invariant isomorphism between the complex vector space of
holomorphic solutions of the wave equation on the forward tube
of flat space-time, and the complex vector space of 
complex-analytic functions of three variables. These are arbitrary,
in that they are not subject to any differential equation.
Eventually, Ward's one-to-one correspondence between complex
space-times with non-vanishing cosmological constant, and
{\it sufficiently small} deformations of flat projective 
twistor space, is presented.

An example of explicit construction of anti-self-dual space-time
is given in Sec. 5, following \cite{Ward78}. This generalization
of Penrose's non-linear graviton 
(\cite{Penr76a}, \cite{Penr76b}) combines
two-spinor techniques and twistor theory in a way very instructive
for beginning graduate students. However, it appears necessary
to go beyond anti-self-dual space-times, since they are only a
particular class of (complex) space-times, and they do not
enable one to recover the full physical content of (complex)
general relativity. This implies going beyond the original
twistor theory, since the three-complex-dimensional space of
$\alpha$-surfaces only exists in anti-self-dual space-times.
After a brief review of alternative ideas, attention is focused
on the recent attempt by Roger Penrose to define twistors as
{\it charges} for massless spin-${3\over 2}$ fields. Such an
approach has been considered since a vanishing Ricci tensor 
provides the consistency condition for the existence and
propagation of massless helicity-${3\over 2}$ fields in
curved space-time. Moreover, in Minkowski space-time the space 
of charges for such fields is naturally identified with the
corresponding twistor space. The resulting geometric scheme
in the presence of curvature is as follows. First, define a
twistor for Ricci-flat space-time. Second, characterize the 
resulting twistor space. Third, reconstruct the original Ricci-flat
space-time from such a twistor space. 
One of the main technical difficulties
of the program proposed by Penrose is to obtain a {\it global}
description of the space of potentials for massless
spin-${3\over 2}$ fields. The corresponding {\it local} theory is
instead used, for other purposes, in \cite{Espo95}.

Last, Sec. 6 reviews the Plebanski contributions to complex
general relativity, i.e. heaven spaces and heavenly equations, while
concluding remarks are presented in Sec. 7.

\section{Two-Component Spinor Calculus}

Spinor calculus is presented by relying on
spin-space formalism. 
Given the existence of unprimed and primed spin-space,
one has the isomorphism between such vector spaces and
their duals, realized by a symplectic form. Moreover, for
Lorentzian metrics, complex conjugation is the 
(anti-)isomorphism between unprimed and primed spin-space.
Finally, for any space-time point, its tangent space is
isomorphic to the tensor product of unprimed and primed 
spin-spaces via the Infeld--van der Waerden symbols.
Hence the correspondence between tensor fields and
spinor fields. Euclidean conjugation in Riemannian
geometries is also discussed in detail.
The Maxwell field strength is written in this language,
and many useful identities are given. The curvature spinors
of general relativity are then constructed explicitly, and
the Petrov classification of space-times 
is obtained in terms of the Weyl
spinor for conformal gravity.

\subsection{Spin-spaces}

Two-component spinor calculus is a powerful tool for
studying classical field theories in four-dimensional
space-time models. Within this framework,
the basic object is spin-space,
a two-dimensional complex vector space $S$ with a
symplectic form $\varepsilon$, i.e. an antisymmetric
complex bilinear form. Unprimed spinor indices 
$A,B,...$ take the values $0,1$ whereas primed spinor
indices $A',B',...$ take the values $0',1'$ since there
are actually two such spaces: unprimed spin-space
$(S,\varepsilon)$ and primed spin-space $(S',\varepsilon')$.
The whole two-spinor calculus in {\it Lorentzian}
four-manifolds relies on three fundamental properties
(\cite{Vebl33}, \cite{Ruse37}, \cite{Penr60},
\cite{PeRi84}, \cite{Espo92}, \cite{Espo94}): 

(i) The isomorphism between $\Bigr(S,\varepsilon_{AB}\Bigr)$ and
its dual $\Bigr(S^{*},\varepsilon^{AB}\Bigr)$. This is provided
by the symplectic form $\varepsilon$, which raises and
lowers indices according to the rules
$$
\varepsilon^{AB} \; \varphi_{B}=\varphi^{A} \; \in \; S,
\eqno (2.1.1)
$$
$$
\varphi^{B} \; \varepsilon_{BA}=\varphi_{A} \; \in \; S^{*}.
\eqno (2.1.2)
$$
Thus, since
$$
\varepsilon_{AB}=\varepsilon^{AB}=\pmatrix {0&1\cr -1&0 \cr},
\eqno (2.1.3)
$$
one finds in components $\varphi^{0}=\varphi_{1},
\varphi^{1}=-\varphi_{0}$. 

Similarly, one has the
isomorphism $\Bigr(S',\varepsilon_{A'B'}\Bigr)
\cong \Bigr((S')^{*},\varepsilon^{A'B'}\Bigr)$, which implies
$$
\varepsilon^{A'B'} \; \varphi_{B'}=\varphi^{A'} \; \in \; S',
\eqno (2.1.4)
$$
$$
\varphi^{B'} \; \varepsilon_{B'A'}=\varphi_{A'} \; \in 
\; (S')^{*},
\eqno (2.1.5)
$$
where
$$
\varepsilon_{A'B'}=\varepsilon^{A'B'}=\pmatrix
{0'&1'\cr -1'&0'\cr}.
\eqno (2.1.6)
$$

(ii) The (anti-)isomorphism between $\Bigr(S,\varepsilon_{AB}\Bigr)$
and $\Bigr(S',\varepsilon_{A'B'}\Bigr)$, called complex conjugation,
and denoted by an overbar. According to a standard convention,
one has
$$
{\overline {\psi^{A}}} \equiv {\overline \psi}^{A'}
\; \in \; S',
\eqno (2.1.7)
$$
$$
{\overline {\psi^{A'}}} \equiv {\overline \psi}^{A}
\; \in \; S.
\eqno (2.1.8)
$$
Thus, complex conjugation maps elements of a spin-space to
elements of the {\it complementary} spin-space. Hence 
some authors say it is an anti-isomorphism.
In components, if $w^{A}$ is thought as
$w^{A}=\pmatrix {\alpha \cr \beta \cr}$, the action of (2.1.7)
leads to
$$
{\overline {w^{A}}} \equiv {\overline w}^{A'}
\equiv \pmatrix {{\overline \alpha} \cr {\overline \beta}\cr},
\eqno (2.1.9)
$$
whereas, if $z^{A'}=\pmatrix {\gamma \cr \delta \cr}$, then
(2.1.8) leads to
$$
{\overline {z^{A'}}} \equiv {\overline z}^{A}
=\pmatrix {{\overline \gamma}\cr {\overline \delta}\cr} .
\eqno (2.1.10)
$$
With our notation, $\overline \alpha$ denotes complex
conjugation of the function $\alpha$, and so on. Note that
the symplectic structure is preserved by complex conjugation,
since ${\overline \varepsilon}_{A'B'}=\varepsilon_{A'B'}$.

(iii) The isomorphism between the tangent space $T$ at a
point of space-time and the tensor product of the 
unprimed spin-space $\Bigr(S,\varepsilon_{AB}\Bigr)$ and the
primed spin-space $\Bigr(S',\varepsilon_{A'B'}\Bigr)$:
$$
T \cong \Bigr(S,\varepsilon_{AB}\Bigr) \otimes
\Bigr(S',\varepsilon_{A'B'}\Bigr).
\eqno (2.1.11)
$$
The Infeld--van der
Waerden symbols $\sigma_{\; \; AA'}^{a}$ and
$\sigma_{a}^{\; \; AA'}$ express this isomorphism, and the
correspondence between a vector $v^{a}$ and a spinor 
$v^{AA'}$ is given by
$$
v^{AA'} \equiv v^{a} \; \sigma_{a}^{\; \; AA'},
\eqno (2.1.12)
$$
$$
v^{a} \equiv v^{AA'} \; \sigma_{\; \; AA'}^{a}.
\eqno (2.1.13)
$$
These mixed spinor-tensor symbols obey the identities
$$
{\overline \sigma}_{a}^{\; \; AA'}=\sigma_{a}^{\; \; AA'},
\eqno (2.1.14)
$$
$$
\sigma_{a}^{\; \; AA'} \; \sigma_{\; \; AA'}^{b}
=\delta_{a}^{\; \; b} ,
\eqno (2.1.15)
$$
$$
\sigma_{a}^{\; \; AA'} \; \sigma_{\; \; BB'}^{a}
=\varepsilon_{B}^{\; \; A} \; \varepsilon_{B'}^{\; \; \; A'} ,
\eqno (2.1.16)
$$
$$
\sigma_{[a}^{\; \; AA'} \; \sigma_{b]A}^{\; \; \; \; \; B'}
=-{i\over 2} \; \varepsilon_{abcd} \; \sigma^{cAA'}
\; \sigma_{\; \; A}^{d \; \; B'}.
\eqno (2.1.17)
$$
Similarly, a one-form $\omega_{a}$ has a spinor equivalent
$$
\omega_{AA'} \equiv \omega_{a} \; \sigma_{\; \; AA'}^{a} ,
\eqno (2.1.18)
$$
whereas the spinor equivalent of the metric is
$$
\eta_{ab} \; \sigma_{\; \; AA'}^{a}
\; \sigma_{\; \; BB'}^{b} \equiv
\varepsilon_{AB} \; \varepsilon_{A'B'} .
\eqno (2.1.19)
$$
In particular, in Minkowski space-time, the above equations
enable one to write down a coordinate system
in $2 \times 2$ matrix form
$$
x^{AA'}={1\over \sqrt{2}}
\pmatrix {{x^{0}+x^{3}}&{x^{1}-ix^{2}}\cr
{x^{1}+ix^{2}}&{x^{0}-x^{3}}\cr}.
\eqno (2.1.20)
$$
More precisely, in a (curved) space-time, one should write the
following equation to obtain the spinor equivalent of a vector:
$$
u^{AA'}=u^{a} \; e_{a}^{\; {\hat c}} \; 
\sigma_{{\hat c}}^{\; \; \; AA'} ,
$$
where $e_{a}^{\; {\hat c}}$ is a standard notation for the tetrad,
and $e_{a}^{\; {\hat c}}\sigma_{{\hat c}}^{\; \; \; AA'} \equiv
e_{a}^{\; AA'}$ is called the {\it soldering form}. This is, by
construction, a spinor-valued one-form, which encodes the relevant
information about the metric $g$, because $g_{ab}=e_{a}^{\; {\hat c}}
e_{b}^{\; {\hat d}} \eta_{{\hat c}{\hat d}}$, $\eta$ being the
Minkowskian metric of the so-called ``internal space".

In the Lorentzian-signature case, the Maxwell 
two-form $F \equiv F_{ab}dx^{a} \wedge dx^{b}$ can be
written spinorially \cite{WaWe90} as
$$
F_{AA'BB'}={1\over 2}\Bigr(F_{AA'BB'}-F_{BB'AA'}\Bigr)
=\varphi_{AB} \; \varepsilon_{A'B'}
+\varphi_{A'B'} \; \varepsilon_{AB} ,
\eqno (2.1.21)
$$
where
$$
\varphi_{AB} \equiv {1\over 2} 
F_{AC'B}^{\; \; \; \; \; \; \; \; \; C'}
=\varphi_{(AB)} ,
\eqno (2.1.22)
$$
$$
\varphi_{A'B'} \equiv {1\over 2}
F_{CB' \; \; A'}^{\; \; \; \; \; \; C}
=\varphi_{(A'B')} .
\eqno (2.1.23)
$$
These formulae are obtained by applying the identity
$$
T_{AB}-T_{BA}=\varepsilon_{AB} \; T_{C}^{\; \; C}
\eqno (2.1.24)
$$
to express ${1\over 2}\Bigr(F_{AA'BB'}-F_{AB'BA'}\Bigr)$
and ${1\over 2}\Bigr(F_{AB'BA'}-F_{BB'AA'}\Bigr)$.
Note also that round brackets $(AB)$ denote (as usual)
symmetrization over the spinor indices $A$ and $B$, and that
the antisymmetric part of $\varphi_{AB}$ vanishes by virtue
of the antisymmetry of $F_{ab}$, since \cite{WaWe90}
$\varphi_{[AB]}={1\over 4}\varepsilon_{AB} \;
F_{CC'}^{\; \; \; \; \; CC'}={1\over 2}\varepsilon_{AB}
\; \eta^{cd} \; F_{cd}=0$. Last but not least, in the 
Lorentzian case
$$
{\overline {\varphi_{AB}}} \equiv {\overline \varphi}_{A'B'}
=\varphi_{A'B'}.
\eqno (2.1.25)
$$
The symmetric spinor fields $\varphi_{AB}$ and
$\varphi_{A'B'}$ are the anti-self-dual and self-dual parts
of the curvature two-form, respectively.

Similarly, the Weyl curvature $C_{\; \; bcd}^{a}$, i.e. the
part of the Riemann curvature tensor invariant under conformal
rescalings of the metric, may be expressed spinorially,
omitting soldering forms 
for simplicity of notation, as
$$
C_{abcd}=\psi_{ABCD} \; \varepsilon_{A'B'} \;
\varepsilon_{C'D'}
+{\overline \psi}_{A'B'C'D'} \;
\varepsilon_{AB} \; \varepsilon_{CD}.
\eqno (2.1.26)
$$

In Hamiltonian gravity, two-component spinors lead to
a considerable simplification of calculations. On denoting
by $n^{\mu}$ the future-pointing unit timelike normal to a
spacelike three-surface, its spinor version obeys the relations
$$
n_{AA'} \; e_{\; \; \; \; \; i}^{AA'}=0,
\eqno (2.1.27)
$$
$$
n_{AA'} \; n^{AA'}=1,
\eqno (2.1.28)
$$
where $e_{\; \; \; \; \; \mu}^{AA'} \equiv e_{\; \; \mu}^{a}
\; \sigma_{a}^{\; \; AA'}$ is the two-spinor version of the tetrad,
i.e. the soldering form introduced before. 
Denoting by $h$ the induced metric on the three-surface, other
useful relations are \cite{Espo94}
$$
h_{ij}=-e_{AA'i} \; e_{\; \; \; \; \; j}^{AA'} ,
\eqno (2.1.29)
$$
$$
e_{\; \; \; \; \; 0}^{AA'}=N \; n^{AA'}
+N^{i} \; e_{\; \; \; \; \; i}^{AA'} ,
\eqno (2.1.30)
$$
$$
n_{AA'} \; n^{BA'}={1\over 2} \varepsilon_{A}^{\; \; B} ,
\eqno (2.1.31)
$$
$$
n_{AA'} \; n^{AB'}={1\over 2} \varepsilon_{A'}^{\; \; \; B'} ,
\eqno (2.1.32)
$$
$$
n_{[EB'} \; n_{A]A'}={1\over 4}\varepsilon_{EA} \;
\varepsilon_{B'A'} ,
\eqno (2.1.33)
$$
$$
e_{AA'j} \; e_{\; \; \; \; \; k}^{AB'}
=-{1\over 2}h_{jk} \; \varepsilon_{A'}^{\; \; \; B'}
-i \varepsilon_{jkl}\sqrt{{\rm det} \; h} \;
n_{AA'} \; e^{AB'l} .
\eqno (2.1.34)
$$
In Eq. (2.1.30), $N$ and $N^{i}$ are the lapse and shift
functions respectively \cite{Espo94}.

To obtain the space-time curvature, we first need to define
the spinor covariant derivative $\nabla_{AA'}$. 
If $\theta,\phi,\psi$ are spinor fields, $\nabla_{AA'}$
is a map such that (\cite{PeRi84}, \cite{Stew91})
\vskip 0.3cm
\noindent
(1) $\nabla_{AA'}(\theta+\phi)=\nabla_{AA'}\theta
+\nabla_{AA'}\phi$ (i.e. linearity).
\vskip 0.3cm
\noindent
(2) $\nabla_{AA'}(\theta \psi)=\Bigr(\nabla_{AA'}\theta\Bigr)\psi
+\theta \Bigr(\nabla_{AA'}\psi\Bigr)$ (i.e. Leibniz rule).
\vskip 0.3cm
\noindent
(3) $\psi=\nabla_{AA'}\theta$ implies
${\overline \psi}=\nabla_{AA'}{\overline \theta}$
(i.e. reality condition).
\vskip 0.3cm
\noindent
(4) $\nabla_{AA'}\varepsilon_{BC}=\nabla_{AA'}\varepsilon^{BC}=0$,
i.e. the symplectic form may be used to raise or lower indices
within spinor expressions acted upon by $\nabla_{AA'}$, in
addition to the usual metricity condition
$\nabla g=0$, which involves instead the product of two
$\varepsilon$-symbols. 
\vskip 0.3cm
\noindent
(5) $\nabla_{AA'}$ commutes with any index substitution
not involving $A,A'$.
\vskip 0.3cm
\noindent
(6) For any function $f$, one finds
$\Bigr(\nabla_{a}\nabla_{b}-\nabla_{b}\nabla_{a}\Bigr)f
=2S_{ab}^{\; \; \; c} \; \nabla_{c}f$, where 
$S_{ab}^{\; \; \; c}$ is the torsion tensor.
\vskip 0.3cm
\noindent
(7) For any derivation $D$ acting on spinor fields, a spinor
field $\xi^{AA'}$ exists such that $D \psi=\xi^{AA'}
\; \nabla_{AA'} \psi, \forall \psi$.
\vskip 0.3cm
\noindent
As proved in \cite{PeRi84}, such a spinor covariant
derivative exists and is unique.

If Lorentzian space-time is replaced by a complex or
real Riemannian four-manifold, an important modification
should be made, since the (anti-)isomorphism between
unprimed and primed spin-space no longer exists. This
means that primed spinors can no longer be regarded as
complex conjugates of unprimed spinors, or viceversa,
as in (2.1.7) and (2.1.8). In particular, Eqs. (2.1.21)
and (2.1.26) should be re-written as
$$
F_{AA'BB'}=\varphi_{AB} \; \varepsilon_{A'B'}
+{\widetilde \varphi}_{A'B'} \; \varepsilon_{AB} ,
\eqno (2.1.35)
$$
$$
C_{abcd}=\psi_{ABCD} \; \varepsilon_{A'B'}
\; \varepsilon_{C'D'}
+{\widetilde \psi}_{A'B'C'D'} \; 
\varepsilon_{AB} \; \varepsilon_{CD} .
\eqno (2.1.36)
$$
With our notation, $\varphi_{AB},{\widetilde \varphi}_{A'B'}$,
as well as $\psi_{ABCD},{\widetilde \psi}_{A'B'C'D'}$
are {\it completely independent} symmetric spinor fields,
not related by any conjugation.

Indeed, a conjugation can still be defined in the real
Riemannian case, but it no longer 
relates $\Bigr(S,\varepsilon_{AB}\Bigr)$
to $\Bigr(S',\varepsilon_{A'B'}\Bigr)$. It is instead an
anti-involutory operation which maps elements of a spin-space
(either unprimed or primed) to elements of the {\it same}
spin-space. By anti-involutory we mean that, when applied twice
to a spinor with an odd number of indices, 
it yields the same spinor with the opposite
sign, i.e. its square is minus the identity, whereas the square
of complex conjugation as defined in (2.1.9) and (2.1.10) equals
the identity. Following \cite{Wood85} and \cite{Espo94},
Euclidean conjugation, denoted by a {\it dagger}, is defined by
$$
{\Bigr(w^{A}\Bigr)}^{\dagger} \equiv 
\pmatrix {{\overline \beta}\cr -{\overline \alpha}\cr} ,
\eqno (2.1.37)
$$
$$
{\Bigr(z^{A'}\Bigr)}^{\dagger} \equiv
\pmatrix {-{\overline \delta}\cr {\overline \gamma}\cr} .
\eqno (2.1.38)
$$
This means that, in flat Euclidean four-space, a unit
$2 \times 2$ matrix $\delta_{BA'}$ exists such that
$$
{\Bigr(w^{A}\Bigr)}^{\dagger} \equiv
\varepsilon^{AB} \; \delta_{BA'} \;
{\overline w}^{A'} .
\eqno (2.1.39)
$$
We are here using the freedom to regard $w^{A}$ either as an
$SL(2,C)$ spinor for which complex conjugation can be defined,
or as an $SU(2)$ spinor for which Euclidean conjugation is
instead available. The soldering forms for $SU(2)$ spinors
only involve spinor indices of the same spin-space, i.e.
${\widetilde e}_{i}^{\; \; AB}$ and 
${\widetilde e}_{i}^{\; \; A'B'}$.
More precisely, denoting by $E_{a}^{i}$
a real {\it triad}, where $i=1,2,3$, and by 
$\tau_{\; \; A}^{a \; \; \; B}$ the three Pauli matrices,
the $SU(2)$ soldering forms are defined by 
$$
{\widetilde e}_{\; \; A}^{j \; \; \; B} \equiv
-{i \over \sqrt{2}} \; E_{a}^{j} \; \tau_{\; \; A}^{a \; \; \; B}.
\eqno (2.1.40)
$$
The soldering form in (2.1.40)
provides an isomorphism between the three-real-dimensional tangent
space at each point of $\Sigma$, and the three-real-dimensional
vector space of $2 \times 2$ trace-free Hermitian matrices.
The Riemannian three-metric on $\Sigma$ is then given by
$$
h^{ij}=-{\widetilde e}_{\; \; A}^{i \; \; \; B} \; 
{\widetilde e}_{\; \; B}^{j \; \; \; A}.
\eqno (2.1.41)
$$

\subsection{Curvature in general relativity}

At this stage, following \cite{PeRi84}, we
want to derive the spinorial form of the Riemann curvature
tensor in a Lorentzian space-time with vanishing torsion, 
starting from the well-known symmetries of Riemann. In
agreement with the abstract-index translation of tensors
into spinors, soldering forms will be omitted in the
resulting equations.

Since $R_{abcd}=-R_{bacd}$ we may write
$$ 
R_{abcd}=R_{AA'BB'CC'DD'}
={1\over 2} 
R_{AF'B \; \; \; cd}^{\; \; \; \; \; \; \; \; \; F'}
\; \varepsilon_{A'B'} 
+{1\over 2} 
R_{FA' \; \; B'cd}^{\; \; \; \; \; \; F}
\; \varepsilon_{AB} .
\eqno (2.2.1) 
$$
Moreover, on defining
$$
X_{ABCD} \equiv {1\over 4} 
R_{AF'B \; \; \; CL'D}^{\; \; 
\; \; \; \; \; \; \; F' \; \; \; \; \; \; \; \; \; \; L'} ,
\eqno (2.2.2)
$$
$$
\Phi_{ABC'D'} \equiv {1\over 4}
R_{AF'B \; \; \; LC' \; \; D'}^{\; \; 
\; \; \; \; \; \; \; F' \; \; \; \; \; \; L} ,
\eqno (2.2.3)
$$
the anti-symmetry in $cd$ leads to
$$ 
R_{abcd}=X_{ABCD} \; \varepsilon_{A'B'} \; \varepsilon_{C'D'}
+\Phi_{ABC'D'} \; \varepsilon_{A'B'} \; \varepsilon_{CD}
$$
$$
+{\overline \Phi}_{A'B'CD} \; \varepsilon_{AB} \; \varepsilon_{C'D'}
+{\overline X}_{A'B'C'D'} \; \varepsilon_{AB} \;
\varepsilon_{CD} .
\eqno (2.2.4) 
$$
According to a standard terminology, the spinors (2.2.2) and (2.2.3)
are called the {\it curvature spinors}. In the light of the
(anti-)symmetries of $R_{abcd}$, they have the following
properties:
$$
X_{ABCD}=X_{(AB)(CD)} ,
\eqno (2.2.5)
$$
$$
\Phi_{ABC'D'}=\Phi_{(AB)(C'D')} ,
\eqno (2.2.6)
$$
$$
X_{ABCD}=X_{CDAB} ,
\eqno (2.2.7)
$$
$$
{\overline \Phi}_{ABC'D'}=\Phi_{ABC'D'} .
\eqno (2.2.8)
$$
Remarkably, Eqs. (2.2.6) and (2.2.8) imply that
$\Phi_{AA'BB'}$ corresponds to a trace-free and real tensor:
$$
\Phi_{a}^{\; \; a}=0 , \; 
\Phi_{AA'BB'}=\Phi_{ab}={\overline \Phi}_{ab} .
\eqno (2.2.9)
$$
Moreover, from Eqs. (2.2.5) and (2.2.7) one obtains
$$
X_{A(BC)}^{\; \; \; \; \; \; \; \; \; A}=0 .
\eqno (2.2.10)
$$
Three duals of $R_{abcd}$ exist which are very useful and are
defined as follows:
$$
R_{\; \; abcd}^{*} \equiv {1\over 2}
\varepsilon_{cd}^{\; \; \; \; pq} \; R_{abpq}
=i \; R_{AA'BB'CD'DC'} ,
\eqno (2.2.11)
$$
$$
{ }^{*}R_{abcd} \equiv {1\over 2} 
\varepsilon_{ab}^{\; \; \; \; pq} \; R_{pqcd}
=i \; R_{AB'BA'CC'DD'} ,
\eqno (2.2.12)
$$
$$
{ }^{*}R_{\; \; abcd}^{*} \equiv {1\over 4}
\varepsilon_{ab}^{\; \; \; \; pq} \;
\varepsilon_{cd}^{\; \; \; \; rs} \;
R_{pqrs} = -R_{AB'BA'CD'DC'} .
\eqno (2.2.13)
$$
For example, in terms of the dual (2.2.11), the familiar equation
$R_{a[bcd]}=0$ reads
$$
R_{\; \; ab}^{* \; \; \; \; bc}=0 .
\eqno (2.2.14)
$$
Thus, to derive the spinor form of the cyclic identity, one
can apply (2.2.14) to the equation
$$ 
R_{\; \; abcd}^{*}=-i \; X_{ABCD} \; \varepsilon_{A'B'}
\; \varepsilon_{C'D'}
+i \; \Phi_{ABC'D'} \; \varepsilon_{A'B'} \; \varepsilon_{CD} 
$$
$$
\; \; \; \; \; \; \; \; \; \;  
\; \; \; \; \; \; \; \; \; \;  
-i \; {\overline \Phi}_{A'B'CD} \; \varepsilon_{AB} \;
\varepsilon_{C'D'}
+i \; {\overline X}_{A'B'C'D'} \; \varepsilon_{AB}
\; \varepsilon_{CD} .
\eqno (2.2.15) 
$$
By virtue of (2.2.6) and (2.2.8) one thus finds
$$
X_{AB \; \; \; C}^{\; \; \; \; \; B} \; \varepsilon_{A'C'}
={\overline X}_{A'B' \; \; \; C'}^{\; \; \; \; \; \; \; B'}
\; \varepsilon_{AC} ,
\eqno (2.2.16)
$$
which implies, on defining
$$
\Lambda \equiv {1\over 6} X_{AB}^{\; \; \; \; \; AB} ,
\eqno (2.2.17)
$$
the reality condition
$$
\Lambda={\overline \Lambda} .
\eqno (2.2.18)
$$

Equation (2.2.1) enables one to express the Ricci tensor
$R_{ab} \equiv R_{acb}^{\; \; \; \; \; c}$ in spinor form as
$$
R_{ab}=6\Lambda \; \varepsilon_{AB} \; \varepsilon_{A'B'}
-2\Phi_{ABA'B'} .
\eqno (2.2.19)
$$
Thus, the resulting scalar curvature, trace-free part of Ricci
and Einstein tensor are
$$
R=24 \Lambda ,
\eqno (2.2.20)
$$
$$
R_{ab}-{1\over 4} R \; g_{ab}=-2 \Phi_{ab}
=-2 \Phi_{ABA'B'} ,
\eqno (2.2.21)
$$
$$
G_{ab}=R_{ab}-{1\over 2} R \; g_{ab}
=-6 \Lambda \; \varepsilon_{AB} \; \varepsilon_{A'B'}
-2 \Phi_{ABA'B'} ,
\eqno (2.2.22)
$$
respectively.

We have still to obtain a more suitable form of the Riemann
curvature. For this purpose, following again \cite{PeRi84},
we point out that the curvature spinor
$X_{ABCD}$ can be written as
$$ 
X_{ABCD}={1\over 3} \Bigr(X_{ABCD}+X_{ACDB}+X_{ADBC}\Bigr)
+{1\over 3} \Bigr(X_{ABCD}-X_{ACBD}\Bigr)
$$
$$
+{1\over 3} \Bigr(X_{ABCD}-X_{ADCB}\Bigr)
=X_{(ABCD)}+{1\over 3} \varepsilon_{BC} \;
X_{AF \; \; D}^{\; \; \; \; \; F}
+{1\over 3} \varepsilon_{BD} \;
X_{AFC}^{\; \; \; \; \; \; \; \; F} .
\eqno (2.2.23) 
$$
Since $X_{AFC}^{\; \; \; \; \; \; \; F}=3 \Lambda \;
\varepsilon_{AF}$, Eq. (2.2.23) leads to
$$
X_{ABCD}=\psi_{ABCD}+\Lambda \Bigr(\varepsilon_{AC} \;
\varepsilon_{BD}+\varepsilon_{AD} \; \varepsilon_{BC} \Bigr) ,
\eqno (2.2.24)
$$
where $\psi_{ABCD}$ is the Weyl spinor.

Since $\Lambda={\overline \Lambda}$ from (2.2.18), the
insertion of (2.2.24) into (2.2.4), jointly with the
identity
$$
\varepsilon_{A'B'} \; \varepsilon_{C'D'}
+\varepsilon_{A'D'} \; \varepsilon_{B'C'}
-\varepsilon_{A'C'} \; \varepsilon_{B'D'}=0 ,
\eqno (2.2.25)
$$
yields the desired decomposition of the Riemann curvature as
$$ 
R_{abcd}=\psi_{ABCD} \; \varepsilon_{A'B'} \; \varepsilon_{C'D'}
+{\overline \psi}_{A'B'C'D'} \; \varepsilon_{AB} \;
\varepsilon_{CD} 
$$
$$
\; \; \; \; \; \; \; \; \; \; 
+\Phi_{ABC'D'} \; \varepsilon_{A'B'} \; \varepsilon_{CD}
+{\overline \Phi}_{A'B'CD} \; \varepsilon_{AB} \;
\varepsilon_{C'D'}
$$
$$
\; \; \; \; \; \; \; \; \; \; \; \; \; \; \; \; 
\; \; \; \; \; \; \; \; \; \; \; \; \; \; \; \; 
+2 \Lambda \Bigr(\varepsilon_{AC} \; \varepsilon_{BD}
\; \varepsilon_{A'C'} \; \varepsilon_{B'D'}
- \varepsilon_{AD} \; \varepsilon_{BC} \; \varepsilon_{A'D'}
\; \varepsilon_{B'C'} \Bigr) .
\eqno (2.2.26) 
$$
With this standard notation, the conformally invariant part of
the curvature takes the form $C_{abcd}={ }^{(-)}C_{abcd}
+{ }^{(+)}C_{abcd}$, where
$$
{ }^{(-)}C_{abcd} \equiv \psi_{ABCD} \; \varepsilon_{A'B'} \;
\varepsilon_{C'D'} ,
\eqno (2.2.27)
$$
$$
{ }^{(+)}C_{abcd} \equiv {\overline \psi}_{A'B'C'D'}
\; \varepsilon_{AB} \; \varepsilon_{CD} ,
\eqno (2.2.28)
$$
are the anti-self-dual and self-dual Weyl tensors,
respectively.

\subsection{Petrov classification}

Since the Weyl spinor is totally symmetric, we may use a
well known result of two-spinor calculus, according to
which, if $\Omega_{AB...L}$ is totally symmetric, then there
exist univalent spinors $\alpha_{A},\beta_{B},...,\gamma_{L}$
such that \cite{Stew91}
$$
\Omega_{AB...L}=\alpha_{(A} \; \beta_{B} ... \gamma_{L)} ,
\eqno (2.3.1)
$$
where $\alpha,...,\gamma$ are called the {\it principal spinors} of
$\Omega$, and the corresponding real null vectors are called the
{\it principal null directions} of $\Omega$. In the case of the Weyl
spinor, such a theorem implies that
$$
\psi_{ABCD}=\alpha_{(A} \; \beta_{B} \; \gamma_{C} \;
\delta_{D)} .
\eqno (2.3.2)
$$
The corresponding space-times can be classified as follows
\cite{Stew91}.
\vskip 0.3cm
\noindent
(1) {\it Type I}. Four distinct principal null directions.
Hence the name algebraically general.
\vskip 0.3cm
\noindent
(2) {\it Type II}. Two directions coincide. Hence the name
algebraically special.
\vskip 0.3cm
\noindent
(3) {\it Type D}. Two different pairs of repeated principal
null directions exist.
\vskip 0.3cm
\noindent
(4) {\it Type III}. Three principal null directions
coincide.
\vskip 0.3cm
\noindent
(5) {\it Type N}. All four principal null directions
coincide.
\vskip 0.3cm
\noindent
Such a classification is the Petrov classification, and it
provides a relevant example of the superiority of the
two-spinor formalism in four space-time dimensions, since the
alternative ways to obtain it are far more complicated.

Within this framework (as well as in Sec. 3) we
need to know that $\psi_{ABCD}$ has two scalar invariants:
$$
I \equiv \psi_{ABCD} \; \psi^{ABCD} ,
\eqno (2.3.3)
$$
$$
J \equiv \psi_{AB}^{\; \; \; \; \; CD} \;
\psi_{CD}^{\; \; \; \; \; EF} \;
\psi_{EF}^{\; \; \; \; \; AB} .
\eqno (2.3.4)
$$
Type-II space-times are such that $I^{3}=6J^{2}$,
while in type-III space-times $I=J=0$. Moreover, type-D
space-times are characterized by the condition
$$
\psi_{PQR(A} \; \psi_{BC}^{\; \; \; \; \; PQ} \;
\psi_{\; \; DEF)}^{R}=0 ,
\eqno (2.3.5)
$$
while in type-N space-times
$$
\psi_{(AB}^{\; \; \; \; \; \; \; EF} \;
\psi_{CD)EF}=0 .
\eqno (2.3.6)
$$
These results, despite their simplicity, are not well known
to many physicists and mathematicians. Hence they have
been included also in this paper, to prepare the ground
for the more advanced topics of the following sections.

\section{Conformal Gravity}

Since twistor theory enables one to reconstruct
the space-time geometry from conformally invariant geometric
objects, it is important to know the basic tools for studying
conformal gravity within the framework of general relativity.
This is achieved by defining and using the Bach \cite{Berg05} and 
Eastwood--Dighton tensors, here presented in two-spinor form
(relying on previous work by Kozameh, Newman and Tod \cite{Koza85}).
After defining $C$-spaces and Einstein spaces, it is shown
that a space-time is conformal to an Einstein space if and only
if some equations involving the Weyl spinor, its covariant
derivatives, and the trace-free part of Ricci are satisfied.
Such a result is then extended to complex Einstein spaces.
The conformal structure of infinity of Minkowski space-time
is eventually introduced.

\subsection{$C$-spaces}

Twistor theory may be viewed as the attempt to describe 
fundamental physics in terms of conformally invariant geometric
objects within a holomorphic framework. Space-time points are no
longer of primary importance, since they only appear as derived
concepts in such a scheme. To understand the following
sections, almost entirely devoted to twistor theory and its
applications, it is therefore necessary to study the main
results of the theory of conformal gravity. They can be
understood by focusing on $C$-spaces, Einstein spaces, complex
space-times and complex Einstein spaces, as we do from now on
in this section.

To study $C$-spaces in a self-consistent way, we begin by recalling
some basic properties of conformal rescalings. By definition, a
{\it conformal rescaling} of the space-time metric $g$ yields the
metric $\widehat g$ as
$$
{\widehat g}_{ab} \equiv e^{2 \omega} \; g_{ab} ,
\eqno (3.1.1)
$$
where $\omega$ is a smooth scalar. Correspondingly, any tensor 
field $T$ of type $(r,s)$ is conformally weighted if
$$
{\widehat T} \equiv e^{k \omega} \; T
\eqno (3.1.2)
$$
for some integer $k$. In particular, conformal invariance of $T$
is achieved if $k=0$.

It is useful to know the transformation rules for covariant
derivatives and Riemann curvature under the rescaling (3.1.1).
For this purpose, defining
$$
F_{\; \; \; ab}^{m} \equiv 2 \delta_{\; \; a}^{m}
\; \nabla_{b} \omega
-g_{ab} \; g^{mn} \; \nabla_{n} \omega ,
\eqno (3.1.3)
$$
one finds
$$
{\widehat \nabla}_{a} \; V_{b}=\nabla_{a} \; V_{b}
-F_{\; \; \; ab}^{m} \; V_{m} ,
\eqno (3.1.4)
$$
where ${\widehat \nabla}_{a}$ denotes covariant differentiation
with respect to the metric $\widehat g$. Hence the Weyl tensor
$C_{abc}^{\; \; \; \; \; d}$, the Ricci tensor 
$R_{ab} \equiv R_{cab}^{\; \; \; \; \; c}$ and the Ricci scalar
transform as
$$
{\widehat C}_{abc}^{\; \; \; \; \; d}=C_{abc}^{\; \; \; \; \; d} ,
\eqno (3.1.5)
$$
$$
{\widehat R}_{ab}=R_{ab}+2 \nabla_{a} \omega_{b}
-2 \omega_{a} \omega_{b}
+g_{ab} \Bigr(2 \omega^{c} \omega_{c}
+\nabla^{c} \omega_{c} \Bigr) ,
\eqno (3.1.6)
$$
$$
{\widehat R}=e^{-2 \omega} \biggr[R+6 \Bigr(\nabla^{c} 
\omega_{c} + \omega^{c} \omega_{c}\Bigr)\biggr] .
\eqno (3.1.7)
$$
With our notation, $\omega_{c} \equiv \nabla_{c} \omega
=\omega_{,c}$.

We are here interested in space-times which are conformal to
$C$-spaces. The latter are a class of space-times such that
$$
{\widehat \nabla}^{f} \; {\widehat C}_{abcf}=0 .
\eqno (3.1.8)
$$
By virtue of (3.1.3) and (3.1.4) one can see that the conformal 
transform of Eq. (3.1.8) is
$$
\nabla^{f} \; C_{abcf}+\omega^{f} \; C_{abcf}=0 .
\eqno (3.1.9)
$$
This is the necessary and sufficient condition for a space-time
to be conformal to a $C$-space. Its two-spinor form is
$$
\nabla^{FA'}\psi_{FBCD}+\omega^{FA'}\psi_{FBCD}=0 .
\eqno (3.1.10)
$$
However, note that only a {\it real} solution $\omega^{FA'}$ of
Eq. (3.1.10) satisfies Eq. (3.1.9). Hence, whenever we use Eq. (3.1.10),
we are also imposing a reality condition \cite{Koza85}. 

On using the invariants defined in (2.3.3) and (2.3.4), one finds
the useful identities
$$
\psi_{ABCD} \; \psi^{ABCE}={1\over 2}I \; \delta_{D}^{\; \; E} ,
\eqno (3.1.11)
$$
$$
\psi_{ABCD} \; \psi_{\; \; \; \; PQ}^{AB} \;
\psi^{PQCE}={1\over 2} J \; \delta_{D}^{\; \; E} .
\eqno (3.1.12)
$$
The idea is now to act with $\psi^{ABCD}$ on the left-hand side
of (3.1.10) and then use (3.1.11) when $I \not = 0$. This leads to
$$
\omega^{AA'}=-{2\over I} \; \psi^{ABCD} \;
\nabla^{FA'} \; \psi_{FBCD} .
\eqno (3.1.13)
$$
By contrast, when $I=0$ but $J \not = 0$, we multiply twice 
Eq. (3.1.10) by the Weyl spinor and use (3.1.12). Hence one
finds
$$
\omega^{AA'}=-{2\over J} \; \psi_{\; \; \; \; \; EF}^{CD}
\; \psi^{EFGA} \; \nabla^{BA'} \; \psi_{BCDG} .
\eqno (3.1.14)
$$
Thus, by virtue of (3.1.13), the reality condition 
$\omega^{AA'}={\overline {\omega^{AA'}}}
={\overline \omega}^{AA'}$ implies
$$
{\overline I} \; \psi^{ABCD} \; \nabla^{FA'} \;
\psi_{FBCD} - I \; {\overline \psi}^{A'B'C'D'}
\; \nabla^{AF'} \; {\overline \psi}_{F'B'C'D'}=0 .
\eqno (3.1.15)
$$
We have thus shown that a space-time is conformally related
to a $C$-space if and only if Eq. (3.1.10) holds for some
vector $\omega^{DD'}=K^{DD'}$, and Eq. (3.1.15) holds as well.

\subsection{3.2 Einstein spaces}

By definition, Einstein spaces are such that their Ricci tensor
is proportional to the metric: $R_{ab}=\lambda \; g_{ab}$. A
space-time is conformal to an Einstein space if and only if
a function $\omega$ exists (see (3.1.1)) such that (cf. (3.1.6))
$$
R_{ab}+2 \nabla_{a} \omega_{b} -2 \omega_{a} \omega_{b}
-{1\over 4} T g_{ab} =0 ,
\eqno (3.2.1)
$$
where
$$
T \equiv R + 2 \nabla^{c} \omega_{c}
-2 \omega^{c} \omega_{c} .
\eqno (3.2.2)
$$
Of course, Eq. (3.2.1) leads to restrictions on the metric. These
are obtained by deriving the corresponding integrability 
conditions. For this purpose, on taking the curl of Eq. (3.2.1)
and using the Bianchi identities, one finds
$$
\nabla^{f} \; C_{abcf}+\omega^{f} \; C_{abcf}=0 ,
$$
which coincides with Eq. (3.1.9). Moreover, acting with $\nabla^{a}$
on Eq. (3.1.9), applying the Leibniz rule, and using again (3.1.9)
to re-express $\nabla^{f} \; C_{abcf}$ as $-\omega^{f} \; C_{abcf}$,
one obtains
$$
\biggr[\nabla^{a}\nabla^{d}+\nabla^{a}\omega^{d}
-\omega^{a}\omega^{d}\biggr]C_{abcd}=0 .
\eqno (3.2.3)
$$
We now re-express $\nabla^{a}\omega^{d}$ from (3.2.1) as
$$
\nabla^{a}\omega^{d}=\omega^{a}\omega^{d}+{1\over 8}T g^{ad}
-{1\over 2}R^{ad} .
\eqno (3.2.4)
$$
Hence Eqs. (3.2.3) and (3.2.4) lead to
$$
\biggr[\nabla^{a}\nabla^{d}-{1\over 2}R^{ad}\biggr]
C_{abcd}=0 .
\eqno (3.2.5)
$$
This calculation only proves that the vanishing of the 
{\it Bach tensor}, defined as
$$
B_{bc} \equiv \nabla^{a}\nabla^{d}C_{abcd}-{1\over 2}R^{ad} \;
C_{abcd} ,
\eqno (3.2.6)
$$
is a {\it necessary} condition for a space-time to be conformal
to an Einstein space (jointly with Eq. (3.1.9)). To prove 
{\it sufficiency} of the condition, we first need the following
Lemma \cite{Koza85}:
\vskip 0.3cm
\noindent
{\bf Lemma 3.2.1} Let $H^{ab}$ be a trace-free symmetric tensor.
Then, providing the scalar invariant $J$ defined in (2.3.4)
does not vanish, the only solution of the equations
$$
C_{abcd} \; H^{ad}=0 ,
\eqno (3.2.7)
$$
$$
C_{\; \; abcd}^{*} \; H^{ad}=0 ,
\eqno (3.2.8)
$$
is $H^{ad}=0$. As shown in Kozameh {\it et al}. (1985), such a Lemma
is best proved by using two-spinor methods. Hence $H_{ab}$
corresponds to the spinor field
$$
H_{AA'BB'}=\phi_{ABA'B'}={\overline \phi}_{(A'B')(AB)} ,
\eqno (3.2.9)
$$
and Eqs. (3.2.7) and (3.2.8) imply that
$$
\psi_{ABCD} \; \phi_{\; \; \; \; \; A'B'}^{CD}=0 .
\eqno (3.2.10)
$$
Note that the extra primed spinor indices $A'B'$ are irrelevant.
Hence we can focus on the simpler eigenvalue equation
$$
\psi_{ABCD} \; \varphi^{CD}= \lambda \; \varphi_{AB} .
\eqno (3.2.11)
$$
The corresponding characteristic equation for $\lambda$ is
$$
-\lambda^{3}+{1\over 2}I \lambda +{\rm det}(\psi)=0 ,
\eqno (3.2.12)
$$
by virtue of (2.3.3). Moreover, the Cayley--Hamilton theorem
enables one to re-write Eq. (3.2.12) as
$$
\psi_{AB}^{\; \; \; \; \; PQ} \;
\psi_{PQ}^{\; \; \; \; \; RS} \;
\psi_{RS}^{\; \; \; \; \; CD}
={1\over 2} I \; \psi_{AB}^{\; \; \; \; \; CD}
+{\rm det}(\psi) \delta_{(A}^{\; \; \; C} \;
\delta_{B)}^{\; \; \; D} ,
\eqno (3.2.13)
$$
and contraction of $AB$ with $CD$ yields
$$
{\rm det}(\psi)={1\over 3}J .
\eqno (3.2.14)
$$
Thus, the only solution of Eq. (3.2.10) is the trivial one unless
$J=0$ \cite{Koza85}.

We are now in a position to prove sufficiency of the conditions
(cf. Eqs. (3.1.9) and (3.2.5))
$$
\nabla^{f} \; C_{abcf}+K^{f} \; C_{abcf}=0 ,
\eqno (3.2.15)
$$
$$
B_{bc}=0 .
\eqno (3.2.16)
$$
Indeed, Eq. (3.2.15) ensures that (3.1.9) is satisfied with
$\omega_{f}=\nabla_{f}\omega$ for some $\omega$. Hence Eq.
(3.2.3) holds. If one now subtracts Eq. (3.2.3) from Eq.
(3.2.16) one finds
$$
C_{abcd}\biggr(R^{ad}+2\nabla^{a}\omega^{d}-2\omega^{a}\omega^{d}
\biggr)=0 .
\eqno (3.2.17)
$$
This is indeed Eq. (3.2.7) of Lemma 3.2.1. To obtain Eq.
(3.2.8), we act with $\nabla^{a}$ on the dual of Eq. (3.1.9).
This leads to
$$
\nabla^{a}\nabla^{d}C_{\; \; abcd}^{*}
+\biggr(\nabla^{a}\omega^{d}-\omega^{a}\omega^{d}\biggr)
C_{\; \; abcd}^{*}=0 .
\eqno (3.2.18)
$$
Following \cite{Koza85}, the gradient of the contracted
Bianchi identity and Ricci identity is then used to derive
the additional equation
$$
\nabla^{a}\nabla^{d}C_{\; \; abcd}^{*}
-{1\over 2}R^{ad} \; C_{\; \; abcd}^{*}=0 .
\eqno (3.2.19)
$$
Subtraction of Eq. (3.2.19) from Eq. (3.2.18) now yields
$$
C_{\; \; abcd}^{*} \biggr(R^{ad}+2\nabla^{a}\omega^{d}
-2\omega^{a}\omega^{d}\biggr)=0 ,
\eqno (3.2.20)
$$
which is the desired form of Eq. (3.2.8).

We have thus completed the proof that (3.2.15) and (3.2.16) are
{\it necessary} and {\it sufficient} conditions for a space-time
to be conformal to an Einstein space. In two-spinor language,
when Einstein's equations are imposed, after a conformal rescaling
the equation for the trace-free part of Ricci becomes
(see Sec. 2.2)
$$
\Phi_{ABA'B'}-\nabla_{BB'}\omega_{AA'}
-\nabla_{BA'}\omega_{AB'}+\omega_{AA'} \; \omega_{BB'}
+\omega_{AB'} \; \omega_{BA'}=0 .
\eqno (3.2.21)
$$
Similarly to the tensorial analysis performed so far, the
spinorial analysis shows that the integrability condition 
for Eq. (3.2.21) is
$$
\nabla^{AA'}\psi_{ABCD}+\omega^{AA'} \; \psi_{ABCD}=0 .
\eqno (3.2.22)
$$
The fundamental theorem of conformal gravity states therefore
that a space-time is conformal to an Einstein space if and
only if \cite{Koza85}
$$
\nabla^{DD'}\psi_{ABCD}+k^{DD'} \; \psi_{ABCD}=0 ,
\eqno (3.2.23)
$$
$$
{\overline I} \; \psi^{ABCD} \; \nabla^{FA'} \;
\psi_{FBCD} - I \; {\overline \psi}^{A'B'C'D'}
\; \nabla^{AF'} \; {\overline \psi}_{F'B'C'D'}=0 ,
\eqno (3.2.24)
$$
$$
B_{AFA'F'} \equiv 2 \biggr(\nabla_{\; \; A'}^{C}
\; \nabla_{\; \; F'}^{D} \; \psi_{AFCD}
+\Phi_{\; \; \; \; \; A'F'}^{CD} \;
\psi_{AFCD}\biggr)=0 .
\eqno (3.2.25)
$$
Note that reality of Eq. (3.2.25) for the Bach spinor is ensured
by the Bianchi identities.

\subsection{Complex space-times}

Since this paper is devoted to complex general relativity and its
applications, it is necessary to extend the theorem expressed by
(3.2.23)--(3.2.25) to complex space-times. For this purpose, we 
find it appropriate to define and discuss such spaces in more detail
in this section. In this respect, we should say that four distinct
geometric objects are necessary to study real general relativity
and complex general relativity, here defined in four-dimensions
(\cite{PeRi86}, \cite{Espo94}).
\vskip 0.3cm
\noindent
(1) {\it Lorentzian} space-time $(M,g_{L})$. 
This is a Hausdorff four-manifold
$M$ jointly with a symmetric, non-degenerate bilinear form $g_{L}$ to
each tangent space with signature $(+,-,-,-)$ (or $(-,+,+,+)$). The
latter is then called a Lorentzian four-metric $g_{L}$.
\vskip 0.3cm
\noindent
(2) {\it Riemannian} four-space $(M,g_{R})$, where $g_{R}$ is a smooth
and {\it positive-definite} section of the bundle of symmetric
bilinear two-forms on $M$. Hence $g_{R}$ has signature $(+,+,+,+)$.
\vskip 0.3cm
\noindent
(3) {\it Complexified} space-time. This manifold originates from a
real-analytic space-time with real-analytic coordinates
$x^{a}$ and real-analytic Lorentzian metric $g_{L}$ by allowing
the coordinates to become complex, and by an holomorphic
extension of the metric coefficients into the complex domain.
In such manifolds the operation of complex conjugation, taking
any point with complexified coordinates $z^{a}$ into the point
with coordinates ${\overline {z^{a}}}$, still exists. Note that,
however, it is not possible to define reality of tensors at
{\it complex points}, since the conjugate tensor lies at the
complex conjugate point, rather than at the original point.
\vskip 0.3cm
\noindent
(4) {\it Complex} space-time. This is a {\it four-complex-dimensional}
complex-Riemannian manifold, and no four-real-dimensional subspace
has been singled out to give it a reality structure \cite{PeRi86}.
In complex space-times no complex conjugation
exists, since such a map is not invariant under holomorphic
coordinate transformations.
\vskip 0.3cm
\noindent
Thus, the complex-conjugate spinors
$\lambda^{A...M}$ and ${\overline \lambda}^{A'...M'}$ of a
Lorentzian space-time are replaced by {\it independent} spinors
$\lambda^{A...M}$ and ${\widetilde \lambda}^{A'...M'}$.
This means that unprimed and primed spin-spaces become 
unrelated to one another.
Moreover, the complex scalars $\phi$ and ${\overline \phi}$ are
replaced by the pair of {\it independent} complex scalars
$\phi$ and $\widetilde \phi$. On the other hand, quantities $X$
that are originally {\it real} yield no new quantities, since the
reality condition $X={\overline X}$ becomes $X={\widetilde X}$.
For example, the covariant derivative operator $\nabla_{a}$ of
Lorentzian space-time yields no new operator
${\widetilde \nabla}_{a}$, since it is originally real. One
should instead regard $\nabla_{a}$ as a complex-holomorphic
operator. The spinors $\psi_{ABCD},\Phi_{ABC'D'}$ and the scalar
$\Lambda$ appearing in the Riemann curvature (see (2.2.26)) have as
counterparts the spinors ${\widetilde \psi}_{A'B'C'D'},
{\widetilde \Phi}_{ABC'D'}$ and the scalar $\widetilde \Lambda$.
However, by virtue of the {\it original} reality conditions in
Lorentzian space-time, one has \cite{PeRi86}
$$
{\widetilde \Phi}_{ABC'D'}=\Phi_{ABC'D'} ,
\eqno (3.3.1)
$$
$$
{\widetilde \Lambda}=\Lambda ,
\eqno (3.3.2)
$$
while the Weyl spinors $\psi_{ABCD}$ and
${\widetilde \psi}_{A'B'C'D'}$ remain independent of each other.
Hence one Weyl spinor may vanish without the other Weyl spinor
having to vanish as well. Correspondingly, a complex space-time
such that ${\widetilde \psi}_{A'B'C'D'}=0$ is called {\it right
conformally flat} or conformally anti-self-dual, whereas if
$\psi_{ABCD}=0$, one deals with a {\it left conformally flat}
or conformally self-dual complex space-time. Moreover, if the
remaining part of the Riemann curvature vanishes as well, i.e.
$\Phi_{ABC'D'}=0$ and $\Lambda=0$, the word {\it conformally}
should be omitted in the terminology described above (cf. 
Sec. 4). Interestingly, in a complex space-time the
principal null directions (cf. Sec. 2.3) of the Weyl spinors
$\psi_{ABCD}$ and ${\widetilde \psi}_{A'B'C'D'}$ are independent
of each other, and one has two independent classification 
schemes at each point.

\subsection{Complex Einstein spaces}

In the light of the previous discussion, the fundamental theorem
of conformal gravity in complex space-times can be stated as
follows \cite{BaMa87}.
\vskip 0.3cm
\noindent
{\bf Theorem 3.4.1} A complex space-time is conformal to a complex
Einstein space if and only if
$$
\nabla^{DD'} \; \psi_{ABCD}+k^{DD'} \; \psi_{ABCD}=0 ,
\eqno (3.4.1)
$$
$$
{\widetilde I} \; \psi^{ABCD} \; \nabla^{FA'} \; \psi_{FBCD}
-I \; {\widetilde \psi}^{A'B'C'D'} \; \nabla^{AF'}
\; {\widetilde \psi}_{F'B'C'D'}=0 ,
\eqno (3.4.2)
$$
$$
B_{AFA'F'} \equiv 2 \biggr(\nabla_{\; \; A'}^{C} \;
\nabla_{\; \; F'}^{D} \; \psi_{AFCD}
+\Phi_{\; \; \; \; \; A'F'}^{CD} \; \psi_{AFCD}\biggr)=0 ,
\eqno (3.4.3)
$$
where $I$ is the complex scalar invariant defined in (2.3.3),
whereas $\widetilde I$ is the independent invariant defined as
$$
{\widetilde I} \equiv {\widetilde \psi}_{A'B'C'D'} \;
{\widetilde \psi}^{A'B'C'D'}.
\eqno (3.4.4)
$$
The left-hand side of Eq. (3.4.2) is called the
{\it Eastwood--Dighton spinor}, and the left-hand side of
Eq. (3.4.3) is the {\it Bach spinor}.

\subsection{Conformal infinity}

To complete our introduction to conformal gravity, we find it
helpful for the reader to outline the construction of
conformal infinity for Minkowski space-time. 
Starting from polar local
coordinates in Minkowski, we first introduce (in $c=1$
units) the retarded coordinate $w \equiv t-r$ and the advanced
coordinate $v \equiv t+r$. To eliminate the resulting cross
term in the local form of the metric, new coordinates
$p$ and $q$ are defined implicitly as \cite{Espo94}
$$
\tan p \equiv v , \;
\tan q \equiv w , \;
p-q \geq 0 .
\eqno (3.5.1)
$$
Hence one finds that a conformal-rescaling factor 
$\omega \equiv (\cos p)(\cos q)$ exists such that, locally,
the metric of Minkowski space-time can be
written as $\omega^{-2}{\widetilde g}$,
where
$$
{\widetilde g} \equiv -dt' \otimes dt'
+\biggr[dr' \otimes dr' +{1\over 4}
(\sin(2r'))^{2} \; \Omega_{2}\biggr],
\eqno (3.5.2)
$$
where $t' \equiv {(p+q)\over 2}, r' \equiv {(p-q)\over 2}$, and
$\Omega_{2}$ is the metric on a unit two-sphere. Although (3.5.2)
is locally identical to the metric of the Einstein static universe,
it is necessary to go beyond a local analysis. This may be
achieved by {\it analytic extension} to the whole of the Einstein
static universe. The original Minkowski space-time is then found
to be conformal to the following region of the Einstein static
universe:
$$
(t'+r') \in ]-\pi,\pi[ , \;
(t'-r') \in ]-\pi,\pi[ , \;
r' \geq 0 .
\eqno (3.5.3)
$$
By definition, the {\it boundary} of the region in (3.5.3) represents
{\it the conformal structure of infinity} of Minkowski space-time.
It consists of two null surfaces and three points, i.e.
\cite{Espo94}
\vskip 0.3cm
\noindent
(i) The null surface ${\rm SCRI}^{-} \equiv
\left \{ t'-r'=q=-{\pi \over 2} \right \}$, i.e. the 
future light cone of the point $r'=0,t'=-{\pi \over 2}$.
\vskip 0.3cm
\noindent
(ii) The null surface ${\rm SCRI}^{+} \equiv 
\left \{t'+r'=p={\pi \over 2} \right \}$, i.e. the past
light cone of the point $r'=0,t'={\pi \over 2}$.
\vskip 0.3cm
\noindent
(iii) Past timelike infinity, i.e. the point
$$
\iota^{-} \equiv \left \{r'=0,t'=-{\pi \over 2} \right \}
\Rightarrow p=q=-{\pi \over 2} .
$$
\vskip 0.3cm
\noindent
(iv) Future timelike infinity, defined as
$$
\iota^{+} \equiv \left \{r'=0,t'={\pi \over 2}
\right \} \Rightarrow p=q={\pi \over 2} .
$$
\vskip 0.3cm
\noindent
(v) Spacelike infinity, i.e. the point
$$
\iota^{0} \equiv \left \{r'={\pi \over 2},t'=0
\right \} \Rightarrow p=-q={\pi \over 2} .
$$
The extension of the SCRI formalism to curved space-times is
an open research problem, but we limit ourselves to the
previous definitions in this section.

\section{Twistor Spaces}

In twistor theory, $\alpha$-planes are the
building blocks of classical field theory in complexified
compactified Minkowski space-time. The $\alpha$-planes are totally
null two-surfaces $S$ in that, if $p$ is any point on $S$,
and if $v$ and $w$ are any two null tangent vectors at 
$p \in S$, the complexified Minkowski metric $\eta$ satisfies
the identity $\eta(v,w)=v_{a}w^{a}=0$. By definition, their
null tangent vectors have the two-component spinor form
$\lambda^{A}\pi^{A'}$, where $\lambda^{A}$ is varying and
$\pi^{A'}$ is fixed. Therefore, the induced metric vanishes
identically since $\eta(v,w)=\Bigr(\lambda^{A}\pi^{A'}\Bigr)
\Bigr(\mu_{A}\pi_{A'}\Bigr)=0=\eta(v,v)=
\Bigr(\lambda^{A}\pi^{A'}\Bigr)
\Bigr(\lambda_{A}\pi_{A'}\Bigr)$. One thus obtains a
conformally invariant characterization of flat space-times.
This definition can be generalized to complex or real
Riemannian space-times with non-vanishing curvature, provided
the Weyl curvature is anti-self-dual. One then finds that
the curved metric $g$ is such that $g(v,w)=0$ on $S$, and
the spinor field $\pi_{A'}$ is covariantly constant on $S$.
The corresponding holomorphic two-surfaces are called 
$\alpha$-surfaces, and they form a three-complex-dimensional
family. Twistor space is the space of all $\alpha$-surfaces,
and depends only on the conformal structure of complex
space-time.

Projective twistor space $PT$ is isomorphic to complex projective
space $CP^{3}$. The correspondence between flat space-time
and twistor space shows that complex $\alpha$-planes correspond
to points in $PT$, and real null geodesics to points in $PN$,
i.e. the space of null twistors. Moreover, a complex
space-time point corresponds to a sphere in $PT$, and a real 
space-time point to a sphere in $PN$. Remarkably, the points
$x$ and $y$ are null-separated if and only if the corresponding
spheres in $PT$ intersect. This is the twistor description of
the light-cone structure of Minkowski space-time.

A conformally invariant isomorphism exists between the complex
vector space of holomorphic solutions of $\cstok{\ }\phi=0$
on the forward tube of flat space-time, and the complex
vector space of arbitrary complex-analytic functions of three
variables, not subject to any differential equation. Moreover,
when curvature is non-vanishing, there is a one-to-one
correspondence between complex space-times with 
anti-self-dual Weyl curvature and scalar curvature
$R=24 \Lambda$, and sufficiently small deformations of flat
projective twistor space $PT$ which preserve a one-form
$\tau$ homogeneous of degree $2$ and a three-form $\rho$ 
homogeneous of degree $4$, with $\tau \wedge d\tau
=2\Lambda \rho$. Thus, to solve the anti-self-dual Einstein
equations, one has to study a geometric problem, i.e.
finding the holomorphic curves in deformed projective
twistor space.

\subsection{$\alpha$-planes in Minkowski space-time}

The $\alpha$-planes provide a geometric definition of twistors
in Minkowski space-time. For this purpose, we first complexify
flat space-time, so that real coordinates 
$\Bigr(x^{0},x^{1},x^{2},x^{3}\Bigr)$ are replaced by complex
coordinates $\Bigr(z^{0},z^{1},z^{2},z^{3}\Bigr)$, and we
obtain a four-dimensional complex vector space equipped with
a non-degenerate complex-bilinear form \cite{WaWe90}
$$
(z,w) \equiv z^{0}w^{0}-z^{1}w^{1}-z^{2}w^{2}-z^{3}w^{3} .
\eqno (4.1.1)
$$
The resulting matrix $z^{AA'}$, which, by construction,
corresponds to
the position vector $z^{a}=\Bigr(z^{0},z^{1},z^{2},z^{3}
\Bigr)$, is no longer Hermitian as in the real case. Moreover,
we compactify such a space by identifying future null infinity
with past null infinity (\cite{Penr74}, \cite{PeRi86},
\cite{Espo94}). The resulting manifold is here denoted by
$CM^{\#}$, following \cite{PeRi86}.

In $CM^{\#}$ with metric $\eta$, we consider two-surfaces
$S$ whose tangent vectors have the two-component spinor
form
$$
v^{a}=\lambda^{A}\pi^{A'} ,
\eqno (4.1.2)
$$
where $\lambda^{A}$ is varying and $\pi^{A'}$ is fixed.
This implies that these tangent vectors are null, since
$\eta(v,v)=v_{a}v^{a}=\Bigr(\lambda^{A}\lambda_{A}\Bigr)
\Bigr(\pi^{A'}\pi_{A'}\Bigr)=0$. Moreover, the induced 
metric on $S$ vanishes identically since any two null
tangent vectors $v^{a}=\lambda^{A}\pi^{A'}$ and
$w^{a}=\mu^{A}\pi^{A'}$ at $p \in S$ are orthogonal:
$$
\eta(v,w)=\Bigr(\lambda^{A}\mu_{A}\Bigr)
\Bigr(\pi^{A'}\pi_{A'}\Bigr)=0 ,
\eqno (4.1.3)
$$
where we have used the property $\pi^{A'}\pi_{A'}=
\varepsilon^{A'B'}\pi_{A'}\pi_{B'}=0$. By virtue of (4.1.3),
the resulting $\alpha$-plane is said to be totally null.
A twistor is then an $\alpha$-plane with constant
$\pi_{A'}$ associated to it. Note that two disjoint families
of totally null two-surfaces exist in $CM^{\#}$, since one
might choose null tangent vectors of the form
$$
u^{a}=\nu^{A}\pi^{A'} ,
\eqno (4.1.4)
$$
where $\nu^{A}$ is fixed and $\pi^{A'}$ is varying. The
resulting two-surfaces are called $\beta$-planes \cite{Penr86}.

Theoretical physicists are sometimes more familiar with a
definition involving the vector space of solutions of the
differential equation
$$
{\cal D}_{A'}^{\; \; (A}\omega^{B)}=0 ,
\eqno (4.1.5)
$$
where ${\cal D}$ is the flat connection, 
and ${\cal D}_{AA'}$ the
corresponding spinor covariant derivative. The general
solution of Eq. (4.1.5) in $CM^{\#}$ takes the form
(\cite{PeRi86}, \cite{Espo94})
$$
\omega^{A}={\Bigr(\omega^{o}\Bigr)}^{A}-i \; x^{AA'}
\pi_{A'}^{o} ,
\eqno (4.1.6)
$$
$$
\pi_{A'}=\pi_{A'}^{o} ,
\eqno (4.1.7)
$$
where $\omega_{A}^{o}$ and $\pi_{A'}^{o}$ are arbitrary
constant spinors, and $x^{AA'}$ is the spinor version of
the position vector with respect to some origin. A twistor
is then {\it represented} by the pair of spinor fields
$\Bigr(\omega^{A},\pi_{A'}\Bigr) \Leftrightarrow Z^{\alpha}$
\cite{Penr75}. The twistor equation (4.1.5) is conformally
invariant. This is proved bearing in mind the spinor form 
of the flat four-metric
$$
\eta_{ab}=\varepsilon_{AB} \; \varepsilon_{A'B'} ,
\eqno (4.1.8)
$$
and making the conformal rescaling
$$
{\widehat \eta}_{ab}=\Omega^{2}\eta_{ab} ,
\eqno (4.1.9)
$$
which implies
$$
{\widehat \varepsilon}_{AB}=\Omega \varepsilon_{AB} , \;
{\widehat \varepsilon}_{A'B'}=\Omega \varepsilon_{A'B'} , \;
{\widehat \varepsilon}^{AB}=\Omega^{-1}\varepsilon^{AB} , \;
{\widehat \varepsilon}^{A'B'}=\Omega^{-1}\varepsilon^{A'B'}.
\eqno (4.1.10)
$$
Thus, defining $T_{a} \equiv 
{\cal D}_{a} \Bigr(\log \Omega \Bigr)$
and choosing ${\widehat \omega}^{B}=\omega^{B}$, one finds
(\cite{PeRi86}, \cite{Espo94})
$$
{\widehat {\cal D}}_{AA'}{\widehat \omega}^{B}
={\cal D}_{AA'}\omega^{B}+\varepsilon_{A}^{\; \; B}
\; T_{CA'}\omega^{C} ,
\eqno (4.1.11)
$$
which implies
$$
{\widehat {\cal D}}_{A'}^{\; \; (A}{\widehat \omega}^{B)}
=\Omega^{-1}{\cal D}_{A'}^{\; \; (A}\omega^{B)} .
\eqno (4.1.12)
$$
Note that the solutions of Eq. (4.1.5) are completely determined
by the four complex components at $\rm O$ of $\omega^{A}$
and $\pi_{A'}$ in a spin-frame at $\rm O$. They are a
four-dimensional vector space over the complex numbers,
called twistor space (\cite{PeRi86}, \cite{Espo94}).

Requiring that $\nu_{A}$ be constant over the
$\beta$-planes implies that 
$\nu^{A}\pi^{A'}{\cal D}_{AA'}\nu_{B}=0$,
for each $\pi^{A'}$, i.e. $\nu^{A}{\cal D}_{AA'}\nu_{B}=0$.
Moreover, a scalar product can be defined between the 
$\omega^{A}$ field and the $\nu_{A}$-scaled $\beta$-plane:
$\omega^{A}\nu_{A}$. Its constancy over the $\beta$-plane
implies that \cite{Penr86}
$$
\nu^{A}\pi^{A'}{\cal D}_{AA'}\Bigr(\omega^{B}\nu_{B}\Bigr)=0 ,
\eqno (4.1.13)
$$
for each $\pi^{A'}$, which leads to
$$
\nu_{A}\nu_{B}\Bigr({\cal D}_{A'}^{\; \; (A}
\omega^{B)}\Bigr)=0 ,
\eqno (4.1.14)
$$
for each $\beta$-plane and hence for each $\nu_{A}$. Thus,
Eq. (4.1.14) becomes the twistor equation (4.1.5). In other
words, it is the twistor concept associated with a
$\beta$-plane which is dual to that associated with a solution
of the twistor equation \cite{Penr86}.

Flat projective twistor space $PT$ can be
thought of as three-dimensional complex projective space
$CP^{3}$ (cf. example E2 in Sec. 1.2). This means that we
take the space $C^{4}$ of complex numbers $\Bigr(z^{0},z^{1},
z^{2},z^{3}\Bigr)$ and factor out by the proportionality relation
$\Bigr(\lambda z^{0},...,\lambda z^{3}\Bigr) \sim
\Bigr(z^{0},...,z^{3}\Bigr)$, with $\lambda \in C - \{0\}$. The
homogeneous coordinates $\Bigr(z^{0},...,z^{3}\Bigr)$ are, in the
case of $PT \cong CP^{3}$, as follows:
$\Bigr(\omega^{0},\omega^{1},\pi_{0'},\pi_{1'}\Bigr) \equiv
\Bigr(\omega^{A},\pi_{A'}\Bigr)$. The $\alpha$-planes defined
in this section can be obtained from the equation (cf. (4.1.6))
$$
\omega^{A}=i \; x^{AA'}\pi_{A'} ,
\eqno (4.1.15)
$$
where $\Bigr(\omega^{A},\pi_{A'}\Bigr)$ is regarded as fixed,
with $\pi_{A'} \not =0$. This means that Eq. (4.1.15), considered
as an equation for $x^{AA'}$, has as its solution a complex
two-plane in $CM^{\#}$, whose tangent vectors take the form
in Eq. (4.1.2), i.e. we have found an $\alpha$-plane. The
$\alpha$-planes are self-dual in that, if $v$ and $u$ are any two
null tangent vectors to an $\alpha$-plane, then 
$F \equiv v \otimes u - u \otimes v$ is a self-dual bivector
since
$$
F^{AA'BB'}=\varepsilon^{AB}\phi^{(A'B')} ,
\eqno (4.1.16)
$$
where $\phi^{(A'B')}=\sigma \pi^{A'}\pi^{B'}$, with 
$\sigma \in C - \{ 0 \}$ (Ward 1981b). Note also that
$\alpha$-planes remain unchanged if we replace 
$\Bigr(\omega^{A},\pi_{A'}\Bigr)$ by
$\Bigr(\lambda \omega^{A},\lambda \pi_{A'}\Bigr)$ with
$\lambda \in C-\{0\}$, and that {\it all} $\alpha$-planes
arise as solutions of Eq. (4.1.15). If real solutions of
such equation exist, this implies that 
$x^{AA'}={\overline x}^{AA'}$. This leads to
$$
\omega^{A}{\overline \pi}_{A}+
{\overline \omega}^{A'}\pi_{A'}=i \; x^{AA'}
\Bigr(\pi_{A'}{\overline \pi}_{A}-\pi_{A'}
{\overline \pi}_{A}\Bigr)=0 ,
\eqno (4.1.17)
$$
where overbars denote complex conjugation in two-spinor
language, defined according to the rules described in
Sec. 2.1. If (4.1.17) holds and $\pi_{A'}
\not =0$, the solution space of Eq. (4.1.15) in real Minkowski
space-time is a null geodesic, and {\it all} null geodesics
arise in this way (Ward 1981b). Moreover, if $\pi_{A'}$
vanishes, the point $\Bigr(\omega^{A},\pi_{A'}\Bigr)
=\Bigr(\omega^{A},0\Bigr)$ can be regarded as an $\alpha$-plane
at infinity in compactified Minkowski space-time. Interestingly,
Eq. (4.1.15) is the two-spinor form of the equation expressing
the incidence property of a point $(t,x,y,z)$ in Minkowski
space-time with the twistor $Z^{\alpha}$, i.e. \cite{Penr81}
$$
\pmatrix {Z^{0}\cr Z^{1}\cr}
={i\over \sqrt{2}}
\pmatrix {t+z & x+iy\cr x-iy & t-z\cr}
\pmatrix {Z^{2}\cr Z^{3}\cr} .
\eqno (4.1.18)
$$
The left-hand side of Eq. (4.1.17) may be then re-interpreted
as the twistor pseudo-norm \cite{Penr81}
$$ 
Z^{\alpha}{\overline Z}_{\alpha}=Z^{0}
{\overline {Z^{2}}}
+Z^{1}{\overline {Z^{3}}}
+Z^{2}{\overline {Z^{0}}}
+Z^{3}{\overline {Z^{1}}}
=\omega^{A}{\overline \pi}_{A}
+\pi_{A'}{\overline \omega}^{A'} ,
\eqno (4.1.19)
$$
by virtue of the property $\Bigr({\overline Z}_{0},
{\overline Z}_{1},{\overline Z}_{2},{\overline Z}_{3}
\Bigr)=\Bigr({\overline {Z^{2}}},
{\overline {Z^{3}}}, {\overline {Z^{0}}},
{\overline {Z^{1}}}\Bigr)$. Such a pseudo-norm makes it possible
to define the {\it top half} $PT^{+}$ of $PT$ by the condition
$Z^{\alpha}{\overline Z}_{\alpha}>0$, and the {\it bottom
half} $PT^{-}$ of $PT$ by the condition
$Z^{\alpha}{\overline Z}_{\alpha}<0$.

So far, we have seen that an $\alpha$-plane corresponds to a
point in $PT$, and null geodesics to points in $PN$, the
space of null twistors. However, we may also interpret (4.1.15)
as an equation where $x^{AA'}$ is fixed, and solve for
$\Bigr(\omega^{A},\pi_{A'}\Bigr)$. Within this framework,
$\pi_{A'}$ remains arbitrary, and $\omega^{A}$ is thus given
by $ix^{AA'}\pi_{A'}$. This yields a complex two-plane, and
factorization by the proportionality relation 
$\Bigr(\lambda \omega^{A},\lambda \pi_{A'}\Bigr) \sim
\Bigr(\omega^{A},\pi_{A'}\Bigr)$ leads to a complex projective
one-space $CP^{1}$, with two-sphere topology. Thus, the fixed
space-time point $x$ determines a Riemann sphere
$L_{x} \cong CP^{1}$ in $PT$. In particular, if $x$ is real,
then $L_{x}$ lies entirely within $PN$, given by those twistors
whose homogeneous coordinates satisfy Eq. (4.1.17). 
To sum up, a complex space-time point corresponds to
a sphere in $PT$, whereas a real space-time point corresponds
to a sphere in $PN$ (\cite{Penr81}, \cite{Ward81b}). 

In Minkowski space-time, two points $p$ and $q$ are 
null-separated if and only if there is a null geodesic 
connecting them. In projective twistor space $PT$, this
implies that the corresponding lines $L_{p}$ and $L_{q}$
intersect, since the intersection point represents the
connecting null geodesic. To conclude this section it may
be now instructive, following \cite{HuTo85}, to
study the relation between null twistors and null geodesics.
Indeed, given the null twistors $X^{\alpha},Y^{\alpha}$
defined by
$$
X^{\alpha} 
\equiv \Bigr(i \; x_{0}^{AC'} \; X_{C'},X_{A'}\Bigr) ,
\eqno (4.1.20)
$$
$$
Y^{\alpha} 
\equiv \Bigr(i \; x_{1}^{AC'} \; Y_{C'},Y_{A'}\Bigr) ,
\eqno (4.1.21)
$$
the corresponding null geodesics are
$$
\gamma_{X}: \; x^{AA'} \equiv x_{0}^{AA'}+\lambda \; 
{\overline X}^{A} \; X^{A'} ,
\eqno (4.1.22)
$$
$$
\gamma_{Y}: \; x^{AA'} \equiv x_{1}^{AA'}+\mu \;
{\overline Y}^{A} \; Y^{A'} .
\eqno (4.1.23)
$$
If these intersect at some point $x_{2}$, one finds
$$
x_{2}^{AA'}=x_{0}^{AA'}+\lambda \; {\overline X}^{A}
\; X^{A'}=x_{1}^{AA'}+\mu \; {\overline Y}^{A} \; Y^{A'} ,
\eqno (4.1.24)
$$
where $\lambda,\mu \in R$. Hence
$$
x_{2}^{AA'} \; {\overline Y}_{A} \; X_{A'}
=x_{0}^{AA'} \; {\overline Y}_{A} \; X_{A'}
=x_{1}^{AA'} \; {\overline Y}_{A} \; X_{A'} ,
\eqno (4.1.25)
$$
by virtue of the identities $X^{A'}X_{A'}=
{\overline Y}^{A} \; {\overline Y}_{A}=0$. Equation (4.1.25)
leads to
$$
X^{\alpha}{\overline Y}_{\alpha}=i\Bigr(x_{0}^{AA'}
\; {\overline Y}_{A} \; X_{A'}
-x_{1}^{AA'} \; {\overline Y}_{A} \; X_{A'}\Bigr)=0 .
\eqno (4.1.26)
$$
Suppose instead we are given Eq. (4.1.26). This implies 
that some real $\lambda$ and $\mu$ exist such that
$$
x_{0}^{AA'}-x_{1}^{AA'}=-\lambda \; {\overline X}^{A}
\; X^{A'}+\mu \; {\overline Y}^{A} \; Y^{A'} ,
\eqno (4.1.27)
$$
where signs on the right-hand side of (4.1.27) have been
suggested by (4.1.24). Note that (4.1.27) only holds if
$X_{A'}Y^{A'} \not =0$, i.e. if $\gamma_{X}$ and 
$\gamma_{Y}$ are not parallel. However, the whole argument can
be generalized to this case as well \cite{HuTo85},
and one finds that in all cases the null geodesics
$\gamma_{X}$ and $\gamma_{Y}$ intersect if and only if
$X^{\alpha} \; {\overline Y}_{\alpha}$ vanishes.

\subsection{$\alpha$-surfaces and twistor geometry}

The $\alpha$-planes defined in Sec. 4.1 can be generalized
to a suitable class of curved complex space-times. 
By a complex space-time $(M,g)$ we mean a
four-dimensional Hausdorff manifold $M$ with holomorphic
metric $g$. Thus, with respect to a holomorphic coordinate basis
$x^{a}$, $g$ is a $4 \times 4$ matrix of holomorphic functions
of $x^{a}$, and its determinant is nowhere-vanishing (\cite{Ward80b},
\cite{WaWe90}). Remarkably, $g$ determines a unique
holomorphic connection $\nabla$, and a holomorphic curvature
tensor $R_{\; bcd}^{a}$. Moreover, the Ricci tensor $R_{ab}$
becomes complex-valued, and the Weyl tensor 
$C_{\; bcd}^{a}$ may be split into {\it independent} holomorphic
tensors, i.e. its self-dual and anti-self-dual parts, respectively.
With our two-spinor notation, one has (see (2.1.36))
$$
C_{abcd}=\psi_{ABCD} \; \varepsilon_{A'B'} \; \varepsilon_{C'D'}
+{\widetilde \psi}_{A'B'C'D'}
\; \varepsilon_{AB} \; \varepsilon_{CD} ,
\eqno (4.2.1)
$$
where $\psi_{ABCD}=\psi_{(ABCD)}, {\widetilde \psi}_{A'B'C'D'}=
{\widetilde \psi}_{(A'B'C'D')}$. The spinors $\psi$ and
$\widetilde \psi$ are the anti-self-dual and self-dual Weyl
spinors, respectively. Following \cite{Penr76a}, \cite{Penr76b}, 
\cite{WaWe90}, complex vacuum space-times such that
$$
{\widetilde \psi}_{A'B'C'D'}=0 , \;
R_{ab}=0 ,
\eqno (4.2.2)
$$
are called {\it right-flat} or {\it anti-self-dual}, whereas
complex vacuum space-times such that
$$
\psi_{ABCD}=0 , \;
R_{ab}=0 ,
\eqno (4.2.3)
$$
are called {\it left-flat} or {\it self-dual}. Note that this
definition only makes sense if space-time is complex or real
Riemannian, since in this case no complex conjugation relates
primed to unprimed spinors (i.e. the corresponding spin-spaces
are no longer anti-isomorphic). Hence, for example, the self-dual
Weyl spinor ${\widetilde \psi}_{A'B'C'D'}$ may vanish without
its anti-self-dual counterpart $\psi_{ABCD}$ having to vanish
as well, as in Eq. (4.2.2), or the converse may hold, as in
Eq. (4.2.3) (see Sec. 1.1 and problem 2.3).

By definition, $\alpha$-surfaces are complex two-surfaces $S$
in a complex space-time $(M,g)$ whose tangent vectors $v$ have
the two-spinor form (4.1.2), where $\lambda^{A}$ is varying,
and $\pi^{A'}$ is a fixed primed spinor field on $S$. From this
definition, the following properties can be derived (cf.
Sec. 4.1).

(i) tangent vectors to $\alpha$-surfaces are null;

(ii) any two null tangent vectors $v$ and $u$ to an
$\alpha$-surface are orthogonal to one another;

(iii) the holomorphic metric $g$ vanishes on $S$ in
that $g(v,u)=g(v,v)=0, \forall v,u$ (cf. (4.1.3)), so
that $\alpha$-surfaces are totally null;

(iv) $\alpha$-surfaces are self-dual, in that 
$F \equiv v \otimes u -u \otimes v$ takes the two-spinor
form (4.1.16);

(v) $\alpha$-surfaces exist in $(M,g)$ if and only if the
self-dual Weyl spinor vanishes, so that $(M,g)$ is
anti-self-dual.
\vskip 0.3cm
\noindent
Note that properties (i)--(iv), here written in a redundant
form for pedagogical reasons, are the same as in the
flat-space-time case, provided we replace the flat metric
$\eta$ with the curved metric $g$. Condition (v), however,
is a peculiarity of curved space-times. 
We here focus on the sufficiency of the condition,
following \cite{WaWe90}.

We want to prove that, if $(M,g)$ is anti-self-dual, it admits
a three-complex-parameter family of self-dual $\alpha$-surfaces.
Indeed, given any point $p \in M$ and a spinor $\mu_{A'}$ at
$p$, one can find a spinor field $\pi_{A'}$ on $M$, satisfying
the equation 
$$
\pi^{A'}\Bigr(\nabla_{AA'}\pi_{B'}\Bigr)=\xi_{A}\pi_{B'} ,
\eqno (4.2.4)
$$
and such that
$$
\pi_{A'}(p)=\mu_{A'}(p) .
\eqno (4.2.5)
$$
Hence $\pi_{A'}$ defines a holomorphic two-dimensional
distribution, spanned by the vector
fields of the form $\lambda^{A}\pi^{A'}$, which is
integrable by virtue of (4.2.4). Thus, in particular, there
exists a self-dual $\alpha$-surface through $p$, with tangent
vectors of the form $\lambda^{A}\mu^{A'}$ at $p$. Since $p$
is arbitrary, this argument may be repeated $\forall p \in M$.
The space $\cal P$ of all self-dual $\alpha$-surfaces in
$(M,g)$ is three-complex-dimensional, and is called twistor
space of $(M,g)$.

\subsection{Geometric theory of partial differential equations}

One of the main results of twistor theory has been a deeper
understanding of the solutions of partial differential
equations of classical field theory. Remarkably, a problem
in analysis becomes a purely geometric problem (\cite{Ward81b},
\cite{WaWe90}). For example, in \cite{Bate04} it was 
shown that the general real-analytic solution of the wave
equation $\cstok{\ }\phi=0$ in Minkowski space-time is
$$
\phi(x,y,z,t)=\int_{-\pi}^{\pi}F(x \cos \theta +y \sin \theta
+iz, y+iz \sin \theta +t \cos \theta, \theta) \; d\theta ,
\eqno (4.3.1)
$$
where $F$ is an arbitrary function of three variables, 
complex-analytic in the first two. Indeed, twistor theory
tells us that $F$ is a function on $PT$. More precisely, let
$f \Bigr(\omega^{A},\pi_{A'}\Bigr)$ be a complex-analytic
function, homogeneous of degree $-2$, i.e. such that
$$
f\Bigr(\lambda \omega^{A},\lambda \pi_{A'}\Bigr)
=\lambda^{-2} f\Bigr(\omega^{A},\pi_{A'}\Bigr) ,
\eqno (4.3.2)
$$
and possibly having singularities \cite{Ward81b}. We now define a
field $\phi(x^{a})$ by
$$
\phi(x^{a}) 
\equiv {1\over 2\pi i}\oint f \Bigr(i \; x^{AA'}
\pi_{A'},\pi_{B'}\Bigr)\pi_{C'} \; d\pi^{C'} ,
\eqno (4.3.3)
$$
where the integral is taken over any closed one-dimensional
contour that avoids the singularities of $f$. Such a field
satisfies the wave equation, and every solution of
$\cstok{\ }\phi=0$ can be obtained in this way. The function
$f$ has been taken to have homogeneity $-2$ since the
corresponding one-form $f \pi_{C'} \; d\pi^{C'}$ has homogeneity
zero and hence is a one-form on projective twistor space $PT$,
or on some subregion of $PT$, since it may have singularities.
The homogeneity is related to the property of $f$ of being 
a free function of three variables. Since $f$ is not defined
on the whole of $PT$, and $\phi$ does not determine $f$ uniquely,
because we can replace $f$ by $f+{\widetilde f}$, where 
$\widetilde f$ is any function such that
$$
\oint {\widetilde f} \pi_{C'} \; d\pi^{C'}=0 ,
\eqno (4.3.4)
$$
we conclude that $f$ is an element of the 
sheaf-cohomology group $H^{1}\Bigr(PT^{+},O(-2)\Bigr)$,
i.e. the complex vector space of arbitrary complex-analytic
functions of three variables, not subject to any differential
equations (\cite{Penr80}, \cite{Ward81b}, \cite{WaWe90}).
Remarkably, a conformally invariant isomorphism exists between
the complex vector space of holomorphic solutions of 
$\cstok{\ }\phi=0$ on the forward tube $CM^{+}$ (i.e. the domain
of definition of positive-frequency fields), and the
sheaf-cohomology group $H^{1}\Bigr(PT^{+},O(-2)\Bigr)$.

It is now instructive to summarize some basic ideas
of sheaf-cohomology theory and its use in
twistor theory, following \cite{Penr80}. For this purpose,
let us begin by recalling how Cech cohomology is obtained. 
We consider a Hausdorff paracompact topological space $X$,
covered with a locally finite system of open sets $U_{i}$.
With respect to this covering, we define a {\it cochain}
with coefficients in an additive Abelian group $G$ (e.g.
$Z, R$ or $C$) in terms of elements $f_{i},f_{ij},f_{ijk}
... \in G$. These elements are assigned to the open sets
$U_{i}$ of the covering, and to their non-empty intersections,
as follows: $f_{i}$ to $U_{i}$, $f_{ij}$ to $U_{i} \cap U_{j}$,
$f_{ijk}$ to $U_{i} \cap U_{j} \cap U_{k}$ and so on. The
elements assigned to non-empty intersections are completely
antisymmetric, so that $f_{i...p}=f_{[i...p]}$. One is thus led
to define
$$
{\rm zero-cochain} \; \alpha 
\equiv \Bigr(f_{1},f_{2},f_{3},...\Bigr) ,
\eqno (4.3.5)
$$
$$
{\rm one-cochain} \; \beta 
\equiv \Bigr(f_{12},f_{23},f_{13},...
\Bigr) ,
\eqno (4.3.6)
$$
$$
{\rm two-cochain} \; \gamma 
\equiv \Bigr(f_{123},f_{124},...\Bigr) ,
\eqno (4.3.7)
$$
and the {\it coboundary operator} $\delta$:
$$
\delta \alpha \equiv \Bigr(f_{2}-f_{1},f_{3}-f_{2},f_{3}-f_{1},
... \Bigr)
\equiv \Bigr(f_{12},f_{23},f_{13},...\Bigr) ,
\eqno (4.3.8)
$$
$$ 
\delta \beta \equiv \Bigr(f_{12}-f_{13}+f_{23},f_{12}-f_{14}
+f_{24},...\Bigr) 
\equiv \Bigr(f_{123},f_{124},...\Bigr) .
\eqno (4.3.9) 
$$
By virtue of (4.3.8) and (4.3.9) one finds $\delta^{2} \alpha =
\delta^{2} \beta =...=0$. {\it Cocycles} $\gamma$ are cochains
such that $\delta \gamma=0$. {\it Coboundaries} are a particular
set of cocycles, i.e. such that $\gamma = \delta \beta$ for some
cochain $\beta$. Of course, {\it all} coboundaries are cocycles,
whereas the converse does not hold. This enables one to define 
the ${\rm p}^{\rm th}$ cohomology group as the quotient space
$$
H_{\Bigr \{U_{i}\Bigr \}}^{p}(X,G) \equiv 
{G_{CC}^{p}/ G_{CB}^{p}} ,
\eqno (4.3.10)
$$
where $G_{CC}^{p}$ is the additive group of p-cocycles,
and $G_{CB}^{p}$ is the additive group of p-coboundaries.
To avoid having a definition which depends on the covering
$\Bigr \{U_{i} \Bigr \}$, one should then take finer and finer
coverings of $X$ and settle on a {\it sufficiently fine}
covering ${\Bigr \{U_{i} \Bigr \}}^{*}$. Following \cite{Penr80},
by this we mean that all the $H^{p} \Bigr(U_{i} \cap
... \cap U_{k},G \Bigr)$ vanish $\forall p >0$. One then
defines
$$
H_{{\Bigr \{U_{i}\Bigr \}}^{*}}^{p}(X,G) \equiv
H^{p}(X,G) .
\eqno (4.3.11)
$$
We always assume such a covering exists, is countable and
locally finite. Note that, rather than thinking of $f_{i}$
as an element of $G$ assigned to $U_{i}$, of $f_{ij}$ as
assigned to $U_{ij}$ and so on, we can think of $f_{i}$ as
a {\it function} defined on $U_{i}$ and taking a constant 
value $\in G$. Similarly, we can think of $f_{ij}$ as a
$G$-valued constant function defined on $U_{i} \cap U_{j}$,
and this implies it is not strictly necessary to assume that
$U_{i} \cap U_{j}$ is non-empty.

The generalization to sheaf cohomology is obtained if we do
not require the functions $f_{i},f_{ij},f_{ijk}...$ to be
constant (there are also cases when the additive group $G$
is allowed to vary from point to point in $X$). The assumption
of main interest is the holomorphic nature of the $f$'s.
A sheaf is so defined that the Cech cohomology previously
defined works as well as before \cite{Penr80}. In other words,
a sheaf $S$ defines an additive group $G_{u}$ for each open set
$U \subset X$. Relevant examples are as follows.

(i) The sheaf $O$ of germs of holomorphic functions on a
complex manifold $X$ is obtained if $G_{u}$ is taken to be
the additive group of all holomorphic functions on $U$.

(ii) Twisted holomorphic functions, i.e. functions whose
values are not complex numbers, but are taken in some complex
line bundle over $X$.

(iii) A particular class of twisted functions is obtained if
$X$ is projective twistor space $PT$ (or $PT^{+}$, or
$PT^{-}$), and the functions studied are holomorphic and
homogeneous of some degree $n$ in the twistor variable, i.e.
$$
f \Bigr (\lambda \omega^{A}, \lambda \pi_{A'} \Bigr)
=\lambda^{n} f \Bigr(\omega^{A},\pi_{A'}\Bigr) .
\eqno (4.3.12)
$$
If $G_{u}$ consists of all such twisted functions on
$U \subset X$, the resulting sheaf, denoted by $O(n)$, is
the sheaf of germs of holomorphic functions twisted by
$n$ on $X$.

(iv) We can also consider vector-bundle-valued functions,
where the vector bundle $B$ is over $X$, and $G_{u}$ consists
of the cross-sections of the portion of $B$ lying above $U$.
\vskip 0.3cm
\noindent
Defining cochains and coboundary operator as before, with
$f_{i} \in G_{U_{i}}$ and so on, we obtain the ${\rm p}^{\rm th}$
cohomology group of $X$, with coefficients in the sheaf $S$, as
the quotient space
$$
H^{p}(X,S) \equiv G^{p}(S)/G_{CB}^{p}(S) ,
\eqno (4.3.13)
$$
where $G^{p}(S)$ is the group of p-cochains with coefficients
in $S$, and $G_{CB}^{p}(S)$ is the group of p-coboundaries
with coefficients in $S$. Again, we take finer and finer
coverings $\Bigr \{U_{i} \Bigr \}$ of $X$, and we settle on a
{\it sufficiently fine} covering. To understand this concept,
we recall the following definitions \cite{Penr80}.
\vskip 0.3cm
\noindent
{\bf Definition 4.3.1} A {\it coherent analytic} sheaf is
locally defined by $n$ holomorphic functions factored out by
a set of $s$ holomorphic relations.
\vskip 0.3cm
\noindent
{\bf Definition 4.3.2} A Stein manifold is a holomorphically
convex open subset of $C^{n}$.
\vskip 0.3cm
\noindent
Thus, we can say that, provided $S$ is a coherent analytic
sheaf, {\it sufficiently fine} means that each of
$U_{i},U_{i}\cap U_{j},U_{i}\cap U_{j} \cap U_{k} ...$
is a Stein manifold. If $X$ is Stein and $S$ is coherent
analytic, then $H^{p}(X,S)=0, \forall p >0$.

We can now consider again the remarks following Eq. (4.3.4), i.e.
the interpretation of twistor functions as elements of
$H^{1}\Bigr(PT^{+},O(-2)\Bigr)$. Let $X$ be a part of $PT$,
e.g. the neighbourhood of a line in $PT$, or the top half
$PT^{+}$, or the closure ${\overline {PT^{+}}}$ of the top
half. We assume $X$ can be covered with two open sets 
$U_{1},U_{2}$ such that every projective line $L$ in $X$ meets
$U_{1} \cap U_{2}$ in an annular region. For us, $U_{1} \cap
U_{2}$ corresponds to the domain of definition of a twistor
function $f(Z^{\alpha})$, homogeneous of degree $n$ in the
twistor $Z^{\alpha}$ (see (4.3.12)). Then 
$f \equiv f_{12} \equiv f_{2}-f_{1}$ is a twisted function
on $U_{1} \cap U_{2}$, and defines a one-cochain $\epsilon$,
with coefficients in $O(n)$, for $X$. By construction
$\delta \epsilon =0$, hence $\epsilon$ is a cocycle. For this
covering, the one-coboundaries are functions of the form
$l_{2}-l_{1}$, where $l_{2}$ is holomorphic on $U_{2}$ and
$l_{1}$ on $U_{1}$. The equivalence between twistor functions
is just the cohomological equivalence between one-cochains
$\epsilon,\epsilon'$ that their difference should be a
coboundary: $\epsilon'-\epsilon=\delta \alpha$, with
$\alpha =\Bigr(l_{1},l_{2}\Bigr)$. This is why we view twistor
functions as defining elements of $H^{1}\Bigr(X,O(n)\Bigr)$.
Indeed, if we try to get finer coverings, we realize it is
often impossible to make $U_{1}$ and $U_{2}$ into Stein manifolds.
However, if $X={\overline {PT^{+}}}$, the covering
$\Bigr \{U_{1},U_{2} \Bigr \}$ by two sets is sufficient for
any analytic, positive-frequency field \cite{Penr80}.

The most striking application of twistor theory to partial
differential equations is perhaps the geometric characterization
of anti-self-dual space-times with a cosmological constant.
For these space-times, the Weyl tensor takes the form
$$
C_{abcd}^{(A.S.D.)}=\psi_{ABCD} \; e_{A'B'} \; e_{C'D'} ,
\eqno (4.3.14)
$$
and the Ricci tensor reads
$$
R_{ab}=-2\Phi_{ab}+6\Lambda g_{ab} .
\eqno (4.3.15)
$$
With our notation, $e_{AB}$ and $e_{A'B'}$ are the curved-space
version of the $\varepsilon$-symbols (denoted again by 
$\varepsilon_{AB}$ and $\varepsilon_{A'B'}$ in Eqs. (2.1.36)
and (4.2.1)), $\Phi_{ab}$ is the trace-free part of Ricci,
$24\Lambda$ is the trace $R=R_{\; \; a}^{a}$ of Ricci
\cite{Ward80b}. The local structure in projective twistor space
which gives information about the metric is a pair of
differential forms: a one-form $\tau$ homogeneous of degree 2
and a three-form $\rho$ homogeneous of degree 4. Basically, 
$\tau$ contains relevant information about $e_{A'B'}$ and 
$\rho$ tells us about $e_{AB}$, hence their knowledge
determines $g_{ab}=e_{AB} \; e_{A'B'}$. The result proved in
\cite{Ward80b} states that a one-to-one correspondence exists 
between sufficiently local anti-self-dual solutions with
scalar curvature $R=24 \Lambda$ and sufficiently small
deformations of flat projective twistor space which preserve 
the one-form $\tau$ and the three-form $\rho$, where
$\tau \wedge d\tau=2\Lambda \rho$. We now describe how to
define the forms $\tau$ and $\rho$, whereas the explicit
construction of a class of anti-self-dual space-times is 
given in Sec. 5.

The geometric framework is twistor space $\cal P$ defined
at the end of Sec. 4.2, i.e. the space of all
$\alpha$-surfaces in $(M,g)$. We take $M$ to be sufficiently
small and convex to ensure that $\cal P$ is a complex
manifold with topology $R^{4} \times S^{2}$, since every point
in an anti-self-dual space-time has such a neighbourhood
\cite{Ward80b}. If $Q$, represented by the pair 
$\Bigr(\alpha^{A},\beta_{A'}\Bigr)$, is any vector in $\cal P$,
then $\tau$ is defined by
$$
\tau(Q) \equiv e^{A'B'} \; \pi_{A'} \; \beta_{B'} .
\eqno (4.3.16)
$$
To make sure $\tau$ is well defined, one has to check that
the right-hand side of (4.3.16) remains covariantly constant
over $\alpha$-surfaces, i.e. is annihilated by the first-order
operator $\lambda^{A}\pi^{A'}\nabla_{AA'}$, since otherwise
$\tau$ does not correspond to a differential form on $\cal P$.
It turns out that $\tau$ is well defined provided the trace-free
part of Ricci vanishes. This is proved using spinor Ricci
identities and the equations of local twistor transport as 
follows \cite{Ward80b}.

Let $v$ be a vector field on the $\alpha$-surface $Z$ such
that $\epsilon v^{a}$ joins $Z$ to the neighbouring 
$\alpha$-surface $Y$. Since $\epsilon v^{a}$ acts as a connecting
vector, the Lie bracket of $v^{a}$ and $\lambda^{B}\pi^{B'}$
vanishes for all $\lambda^{B}$, i.e.
$$
\lambda^{B} \; \pi^{B'} \; \nabla_{BB'} \; v^{AA'}
-v^{BB'} \; \nabla_{BB'} \; \lambda^{A} \; \pi^{A'}=0 .
\eqno (4.3.17)
$$
Thus, after defining
$$
\beta_{A'} \equiv v^{BB'} \; \nabla_{BB'} \; \pi_{A'} ,
\eqno (4.3.18)
$$
one finds
$$
\pi_{A'} \; \lambda^{B} \; \pi^{B'} 
\; \nabla_{BB'} \; v^{AA'}
=\lambda^{A} \; \beta^{A'} \; \pi_{A'} .
\eqno (4.3.19)
$$
If one now applies the torsion-free spinor Ricci identities
(\cite{Penr83}, \cite{PeRi84}), one finds that
the spinor field $\beta_{A'}(x)$ on $Z$ satisfies the
equation
$$
\lambda^{B} \; \pi^{B'} \; \nabla_{BB'} \; \beta_{A'}
=-i \; \lambda^{B} \; \pi^{B'} \; P_{ABA'B'} \; \alpha^{A} ,
\eqno (4.3.20)
$$
where $P_{ab}=\Phi_{ab}-\Lambda g_{ab}$ and $\alpha^{A}
=iv^{AC'} \; \pi_{C'}$. Moreover, Eq. (4.3.19) and the
Leibniz rule imply that
$$
\lambda^{B} \; \pi^{B'} \; \nabla_{BB'} \; \alpha^{A}
=-i \; \lambda^{A} \; \pi^{A'} \; \beta_{A'} ,
\eqno (4.3.21)
$$
since $\pi^{B'}\nabla_{BB'}\pi_{C'}=0$. Equations (4.3.20) and (4.3.21)
are indeed the equations of {\it local twistor transport}, and
Eq. (4.3.20) leads to
$$ 
\lambda^{C}\pi^{C'}\nabla_{CC'}\Bigr(e^{A'B'}\pi_{A'} \; 
\beta_{B'}\Bigr)=e^{A'B'}\pi_{A'}\Bigr(\lambda^{C}\pi^{C'}
\nabla_{CC'}\beta_{B'}\Bigr)
$$
$$
=-i \; \lambda^{B}\pi^{B'} \pi_{C'} \; e^{C'A'}\alpha^{A}
\Bigr(\Phi_{ABA'B'}-\Lambda e_{AB} \; e_{A'B'}\Bigr)
=i \; \lambda^{B}\pi^{A'}\pi^{B'}\alpha^{A}
\Phi_{ABA'B'} ,
\eqno (4.3.22)
$$
since $\pi^{A'}\pi^{B'}e_{A'B'}=0$. Hence, as we said before,
$\tau$ is well defined provided the trace-free part of Ricci
vanishes. Note that, strictly, $\tau$ is a twisted form rather
than a form on $\cal P$, since it is homogeneous of degree 2,
one from $\pi_{A'}$ and one from $\beta_{B'}$. By contrast,
a one-form would be independent of the scaling of $\pi_{A'}$
and $\beta_{B'}$ \cite{Ward80b}.
 
We are now in a position to define the three-form $\rho$, 
homogeneous of degree 4. For this purpose, let us denote by
$Q_{h}$, $h=1,2,3$ three vectors in $\cal P$, represented
by the pairs $\Bigr(\alpha_{h}^{A},\beta_{hA'}\Bigr)$. The
corresponding $\rho(Q_{1},Q_{2},Q_{3})$ is obtained by taking
$$
\rho_{123} \equiv {1\over 2} \Bigr(e^{A'B'}
\pi_{A'} \; \beta_{1B'}\Bigr)
\Bigr(e_{AB} \; \alpha_{2}^{A} \; \alpha_{3}^{B}\Bigr) ,
\eqno (4.3.23)
$$
and then anti-symmetrizing $\rho_{123}$ over $1,2,3$. This
yields
$$
\rho(Q_{1},Q_{2},Q_{3}) \equiv {1\over 6}
\Bigr(\rho_{123}-\rho_{132}+\rho_{231}-\rho_{213}
+\rho_{312}-\rho_{321}\Bigr) .
\eqno (4.3.24)
$$
The reader can check that, by virtue of Eqs. (4.3.20) and (4.3.21),
$\rho$ is well defined, since it is covariantly constant over
$\alpha$-surfaces:
$$
\lambda^{A} \; \pi^{A'} \; 
\nabla_{AA'} \; \rho(Q_{1},Q_{2},Q_{3})=0 .
\eqno (4.3.25)
$$

\section{Penrose Transform for Gravitation}

Deformation theory of complex manifolds
is applied to construct a class of anti-self-dual
solutions of Einstein's vacuum equations, following
the work of Penrose and Ward. The hard part of the
analysis is to find the holomorphic cross-sections of
a deformed complex manifold, and the corresponding
conformal structure of an anti-self-dual
space-time. This calculation is
repeated in detail, using complex analysis and
two-component spinor techniques.

If no assumption about anti-self-duality is made,
twistor theory is by itself insufficient to characterize
geometrically a solution of the full Einstein equations.
After a brief review of alternative ideas based on the 
space of complex null geodesics of complex space-time,
and Einstein-bundle constructions, attention is focused
on the attempt by Penrose to define twistors
as charges for massless spin-${3\over 2}$ fields.
This alternative definition is considered since a vanishing
Ricci tensor provides the consistency condition for the
existence and propagation of massless spin-${3\over 2}$ fields
in curved space-time, whereas in Minkowski space-time the space
of charges for such fields is naturally identified with the
corresponding twistor space.

The two-spinor analysis of the Dirac form of such
fields in Minkowski space-time is carried out in detail by
studying their two potentials with corresponding gauge freedoms.
The Rarita--Schwinger form is also introduced, and self-dual
vacuum Maxwell fields are obtained from massless spin-${3\over 2}$
fields by spin-lowering. In curved space-time, however, 
the local expression of spin-${3\over 2}$ field strengths in terms
of the second of these
potentials is no longer possible, unless one studies
the self-dual Ricci-flat case. Thus, much more work is needed
to characterize geometrically a Ricci-flat (complex) space-time
by using this alternative concept of twistors.

\subsection{Anti-self-dual space-times}

Following \cite{Ward78}, we now use twistor-space techniques
to construct a family of anti-self-dual solutions of
Einstein's vacuum equations. Bearing in mind the
space-time twistor-space correspondence in Minkowskian
geometry described in Sec. 4.1, we take a region
$\cal R$ of $CM^{\#}$, whose corresponding region in $PT$ is
$\widetilde {\cal R}$. Moreover, $\cal N$ is the 
non-projective version of $\widetilde {\cal R}$, which
implies ${\cal N} \subset T \subset C^{4}$. In other
words, as coordinates on $\cal N$ we may use
$\Bigr(\omega^{o},\omega^{1},\pi_{o'},\pi_{1'}\Bigr)$.
The geometrically-oriented reader may like it to know that
three important structures are associated with $\cal N$:

(i) the fibration $\Bigr(\omega^{A},\pi_{A'}\Bigr)
\rightarrow \pi_{A'}$, which implies that $\cal N$ becomes
a bundle over $C^{2}- \{0 \}$;

(ii) the two-form ${1\over 2} d\omega_{A} \wedge 
d\omega^{A}$ on each fibre;

(iii) the projective structure ${\cal N} \rightarrow
{\widetilde {\cal R}}$.
\vskip 0.3cm
\noindent
Deformations of $\cal N$ which preserve this projective
structure correspond to right-flat metrics (see Sec. 4.2)
in $\cal R$. To obtain such deformations, cover $\cal N$ with
two patches $\cal Q$ and ${\widehat {\cal Q}}$. Coordinates on
$\cal Q$ and on ${\widehat {\cal Q}}$ are 
$\Bigr(\omega^{A},\pi_{A'}\Bigr)$ and
$\Bigr({\widehat \omega}^{A}, {\widehat \pi}_{A'}\Bigr)$
respectively. We may now {\it glue} $\cal Q$ and
${\widehat {\cal Q}}$ together according to
$$
{\widehat \omega}^{A}=\omega^{A}
+f^{A}\Bigr(\omega^{B},\pi_{B'}\Bigr) ,
\eqno (5.1.1)
$$
$$
{\widehat \pi}_{A'}=\pi_{A'} ,
\eqno (5.1.2)
$$
where $f^{A}$ is homogeneous of degree 1, holomorphic on
${\cal Q} \bigcap {\widehat {\cal Q}}$, and satisfies
$$
{\rm det} \; \biggr(\varepsilon_{A}^{\; \; B}
+{\partial f^{B} \over \partial \omega^{A}}\biggr)
=1 .
\eqno (5.1.3)
$$
Such a patching process yields a complex manifold 
${\cal N}^{D}$ which is a deformation of $\cal N$. The
corresponding right-flat space-time $\cal G$ is such that
its points correspond to the holomorphic cross-sections of
${\cal N}^{D}$. The hard part of the analysis is indeed to
find these cross-sections, but this can be done explicitly 
for a particular class of patching functions. For this
purpose, we first choose a constant spinor field
$p^{AA'B'}=p^{A(A'B')}$ and a homogeneous holomorphic
function $g(\gamma,\pi_{A'})$ of three complex variables:
$$
g\Bigr(\lambda^{3}\gamma,\lambda \pi_{A'}\Bigr)
=\lambda^{-1}g\Bigr(\gamma,\pi_{A'}\Bigr)
\; \; \forall \lambda \in C - \{0 \} .
\eqno (5.1.4)
$$
This enables one to define the spinor field
$$
p^{A} \equiv p^{AA'B'} \; \pi_{A'} \; \pi_{B'} ,
\eqno (5.1.5)
$$
and the patching function
$$
f^{A} \equiv p^{A} \; g\Bigr(p_{B}\omega^{B},\pi_{B'}\Bigr) ,
\eqno (5.1.6)
$$
and the function
$$
F(x^{a},\pi_{A'}) \equiv g \Bigr(i \; p_{A} \; x^{AC'} \; \pi_{C'},
\pi_{A'}\Bigr) .
\eqno (5.1.7)
$$
Under suitable assumptions on the singularities of $g$, $F$
may turn out to be holomorphic if $x^{a} \in {\cal R}$ and if
the ratio ${\widetilde \pi} \equiv {\pi_{o'} \over \pi_{1'}}
\in ]{1\over 2},{5\over 2}[$. It is also possible to express
$F$ as the difference of two contour integrals after
defining the differential form
$$
\Omega \equiv {\Bigr(2\pi i \rho^{A'}\pi_{A'}\Bigr)}^{-1}
\; F(x^{b},\rho_{B'}) \; \rho_{C'}d\rho^{C'} .
\eqno (5.1.8)
$$
In other words, if $\Gamma$ and $\widehat {\Gamma}$ are closed
contours on the projective $\rho_{A'}$-sphere defined by
$|{\widetilde \rho}|=1$ and
$|{\widetilde \rho}|=2$ respectively, we may define 
the function
$$
h \equiv  \oint_{\Gamma} \Omega ,
\eqno (5.1.9)
$$
holomorphic for ${\widetilde \pi}<2$, and the function
$$
{\widehat h} \equiv \oint_{\widehat \Gamma} \Omega ,
\eqno (5.1.10)
$$
holomorphic for ${\widetilde \pi}>1$. Thus, by virtue of
Cauchy's integral formula, one finds (cf. \cite{Ward78})
$$
F(x^{a},\pi_{A'})={\widehat h}(x^{a},\pi_{A'})
-h(x^{a},\pi_{A'}) .
\eqno (5.1.11)
$$
The basic concepts of sheaf-cohomology presented in Sec.
4.3 are now useful to understand the deep meaning of these
formulae. For any fixed $x^{a}$, $F(x^{a},\pi_{A'})$ 
determines an element of the sheaf-cohomology group 
$H^{1}(P_{1}(C),O(-1))$, where $P_{1}(C)$ is the Riemann
sphere of projective $\pi_{A'}$ spinors and $O(-1)$ is the
sheaf of germs of holomorphic functions of $\pi_{A'}$,
homogeneous of degree $-1$. Since $H^{1}$ vanishes, $F$ is
actually a coboundary. Hence it can be split according to
(5.1.11).

In the subsequent calculations, it will be useful to write
a solution of the Weyl equation $\nabla^{AA'}\psi_{A}=0$
in the form
$$
\psi_{A} \equiv i \; \pi^{A'} \; \nabla_{AA'}h(x^{a},\pi_{C'}).
\eqno (5.1.12)
$$
Moreover, following again \cite{Ward78}, we note that a spinor
field $\xi_{A'}^{\; \; \; B'}(x)$ can be defined by
$$
\xi_{A'}^{\; \; \; B'}\pi_{B'} \equiv
i \; p^{AB'C'} \; \pi_{B'} \; \pi_{C'}
\; \nabla_{AA'}h(x,\pi_{D'}) ,
\eqno (5.1.13)
$$
and that the following identities hold:
$$
i \; p^{AA'B'} \; \pi_{B'} \; \nabla_{AA'}h(x,\pi_{C'})
=\xi \equiv {1\over 2} \xi_{A'}^{\; \; A'} ,
\eqno (5.1.14)
$$
$$
\psi_{A} \; p^{AA'B'}=-\xi^{(A'B')} .
\eqno (5.1.15)
$$

We may now continue the analysis of 
our deformed twistor space ${\cal N}^{D}$,
written in the form (cf. (5.1.1) and (5.1.2))
$$
{\widehat \omega}^{A}=\omega^{A}+p^{A}g
\Bigr(p_{B}\omega^{B},\pi_{B'}\Bigr) ,
\eqno (5.1.16a)
$$
$$
{\widehat \pi}_{A'}=\pi_{A'} .
\eqno (5.1.16b)
$$
In the light of the split (5.1.11), holomorphic sections of
${\cal N}^{D}$ are given by
$$
\omega^{A}(x^{b},\pi_{B'})=i \; x^{AA'} \; \pi_{A'}
+p^{A} \; h(x^{b},\pi_{B'}) 
\; {\rm in} \; {\cal Q} ,
\eqno (5.1.17)
$$
$$
{\widehat \omega}^{A}(x^{b},\pi_{B'})=i \; x^{AA'} \; \pi_{A'}
+p^{A} \; {\widehat h}(x^{b},\pi_{B'})
\; {\rm in} \; {\widehat {\cal Q}} ,
\eqno (5.1.18)
$$
where $x^{b}$ are {\it complex} coordinates on $\cal G$.
The conformal structure of $\cal G$ can be computed as
follows. A vector $U=U^{BB'}\nabla_{BB'}$ at $x^{a} \in
{\cal G}$ may be represented in ${\cal N}^{D}$ by the
displacement
$$
\delta \omega^{A}
=U^{b} \; \nabla_{b} \; \omega^{A}(x^{c},\pi_{C'}) .
\eqno (5.1.19a)
$$
By virtue of (5.1.17), Eq. (5.1.19a) becomes
$$
\delta \omega^{A}=U^{BB'}\Bigr(i \; \varepsilon_{B}^{\; \; A}
\; \pi_{B'}+p^{A} \; \nabla_{BB'}h(x^{c},\pi_{C'})\Bigr) .
\eqno (5.1.19b)
$$
The vector $U$ is null, by definition, if and only if
$$
\delta \omega^{A}(x^{b},\pi_{B'})=0 ,
\eqno (5.1.20)
$$
for some spinor field $\pi_{B'}$. To prove that the solution
of Eq. (5.1.20) exists, one defines (see (5.1.14))
$$
\theta \equiv 1-\xi ,
\eqno (5.1.21)
$$
$$
\Omega_{\; \; \; \; \; AA'}^{BB'} \equiv 
\theta \; \varepsilon_{A}^{\; \; B} \;
\varepsilon_{A'}^{\; \; B'}
-\psi_{A} \; p_{A'}^{\; \; \; BB'} .
\eqno (5.1.22)
$$
We are now aiming to show that the desired solution of
Eq. (5.1.20) is given by 
$$
U^{BB'}=\Omega_{\; \; \; \; \; AA'}^{BB'}
\; \lambda^{A} \; \pi^{A'} .
\eqno (5.1.23)
$$
Indeed, by virtue of (5.1.21)--(5.1.23) one finds
$$
U^{BB'}=(1-\xi)\lambda^{B}\pi^{B'}
-\psi_{A} \; p_{A'}^{\; \; \; BB'}
\; \lambda^{A} \; \pi^{A'} .
\eqno (5.1.24)
$$
Thus, since $\pi^{B'}\pi_{B'}=0$, the calculation of
(5.1.19b) yields
$$ 
\delta \omega^{A}=-\psi_{C} \; \lambda^{C} \; 
\pi^{A'} \Bigr[i \; p_{A'}^{\; \; \; AB'} \; \pi_{B'}
+p_{A'}^{\; \; \; BB'} \; p^{A} \;
\nabla_{BB'}h(x,\pi)\Bigr]
$$
$$
+(1-\xi)\lambda^{B} \; \pi^{B'} \; p^{A}
\; \nabla_{BB'}h(x,\pi) .
\eqno (5.1.25) 
$$
Note that (5.1.12) may be used to re-express the second line
of (5.1.25). This leads to
$$ 
\delta \omega^{A}=-\psi_{C} \; \lambda^{C} \; \Gamma^{A} ,
\eqno (5.1.26)
$$
where
$$ 
\Gamma^{A} \equiv \pi^{A'}\Bigr[i \; p_{A'}^{\; \; \; AB'}
\; \pi_{B'}+p_{A'}^{\; \; \; BB'} \; p^{A} \;
\nabla_{BB'}h(x,\pi)\Bigr]
+i(1-\xi)p^{A}
$$
$$
\; \; \; \; \; \; =-i \; p^{AA'B'} \; \pi_{A'} \; \pi_{B'}+i \; p^{A} 
+p^{A}\Bigr[-p^{BB'A'} \; \pi_{A'} \; \nabla_{BB'}h(x,\pi)
-i\xi \Bigr]
$$
$$
=\Bigr[-i+i+i\xi-i\xi \Bigr]p^{A}=0 ,
\eqno (5.1.27) 
$$
in the light of (5.1.5) and (5.1.14). Hence the solution of Eq. (5.1.20)
is given by (5.1.23).

Such null vectors determine the conformal metric of $\cal G$.
For this purpose, one defines \cite{Ward78}
$$
\nu_{A'}^{\; \; \; B'} \equiv \varepsilon_{A'}^{\; \; \; B'}
-\xi_{A'}^{\; \; \; B'} ,
\eqno (5.1.28)
$$
$$
\Lambda \equiv {\theta \over 2} \; \nu_{A'B'} \; \nu^{A'B'} ,
\eqno (5.1.29)
$$
$$
\Sigma_{BB'}^{\; \; \; \; \; \; CC'} \equiv
\theta^{-1} \; \varepsilon_{B}^{\; \; C} \;
\varepsilon_{B'}^{\; \; \; C'}
+\Lambda^{-1} \; \psi_{B} \; p_{A'}^{\; \; \; CC'}
\; \nu_{B'}^{\; \; \; A'} .
\eqno (5.1.30)
$$
Interestingly, $\Sigma_{b}^{\; \; c}$ is the inverse
of $\Omega_{\; \; a}^{b}$, since
$$
\Omega_{\; \; a}^{b} \; \Sigma_{b}^{\; \; c}
=\delta_{a}^{\; \; c} .
\eqno (5.1.31)
$$
Indeed, after defining
$$
H_{A'}^{\; \; \; CC'} \equiv
p_{A'}^{\; \; \; CC'}-p_{D'}^{\; \; \; CC'}
\; \xi_{A'}^{\; \; \; D'} ,
\eqno (5.1.32)
$$
$$
\Phi_{A'}^{\; \; \; CC'} \equiv
\Bigr[\theta \Lambda^{-1} \; H_{A'}^{\; \; \; CC'}
-\Lambda^{-1} \; p_{A'}^{\; \; \; BB'} \;
\psi_{B} \; H_{B'}^{\; \; \; CC'}
-\theta^{-1} \; p_{A'}^{\; \; \; CC'}\Bigr] ,
\eqno (5.1.33)
$$
a detailed calculation shows that
$$
\Omega_{\; \; \; \; \; AA'}^{BB'} \;
\Sigma_{BB'}^{\; \; \; \; \; \; CC'}
-\varepsilon_{A}^{\; \; C} \;
\varepsilon_{A'}^{\; \; \; C'}
=\psi_{A} \; \Phi_{A'}^{\; \; \; CC'} .
\eqno (5.1.34)
$$
One can now check that the right-hand side of (5.1.34)
vanishes (see problem 5.1). Hence (5.1.31) holds.
For our anti-self-dual space-time $\cal G$, the metric
$g=g_{ab}dx^{a} \otimes dx^{b}$ is such that
$$
g_{ab}=\Xi(x) \; \Sigma_{a}^{\; \; c} \;
\Sigma_{bc} .
\eqno (5.1.35)
$$
Two null vectors $U$ and $V$ at $x \in {\cal G}$ have,
by definition, the form
$$
U^{AA'} \equiv \Omega_{\; \; \; \; \; BB'}^{AA'}
\; \lambda^{B} \; \alpha^{B'} ,
\eqno (5.1.36)
$$
$$
V^{AA'} \equiv \Omega_{\; \; \; \; \; BB'}^{AA'}
\; \chi^{B} \; \beta^{B'} ,
\eqno (5.1.37)
$$
for some spinors $\lambda^{B},\chi^{B},\alpha^{B'},
\beta^{B'}$. In the deformed space ${\cal N}^{D}$, 
$U$ and $V$ correspond to two displacements
$\delta_{1}\omega^{A}$ and $\delta_{2}\omega^{A}$
respectively, as in Eq. (5.1.19{\it b}). If one defines
the corresponding skew-symmetric form
$$
{\cal S}_{\pi}(U,V) \equiv \delta_{1}\omega_{A}
\; \delta_{2}\omega^{A} ,
\eqno (5.1.38)
$$
the metric is given by
$$
g(U,V) \equiv \Bigr(\alpha^{A'} \; \beta_{A'}\Bigr)
{\Bigr(\alpha^{B'} \; \pi_{B'}\Bigr)}^{-1}
{\Bigr(\beta^{C'} \; \pi_{C'}\Bigr)}^{-1}
\; {\cal S}_{\pi}(U,V) .
\eqno (5.1.39)
$$
However, in the light of (5.1.31), (5.1.35)--(5.1.37) one finds
$$
g(U,V) \equiv
g_{ab}U^{a}V^{b}=\Xi(x)\Bigr(\lambda^{A} \; \chi_{A}\Bigr)
\Bigr(\alpha^{A'} \; \beta_{A'}\Bigr) .
\eqno (5.1.40)
$$
By comparison with (5.1.39) this leads to
$$
{\cal S}_{\pi}(U,V)=\Xi(x)\Bigr(\lambda^{A} \; \chi_{A}\Bigr)
\Bigr(\alpha^{B'} \; \pi_{B'}\Bigr)
\Bigr(\beta^{C'} \; \pi_{C'}\Bigr) .
\eqno (5.1.41)
$$
If we now evaluate (5.1.41) with $\beta^{A'}=\alpha^{A'}$,
comparison with the definition (5.1.38) and use of
(5.1.12), (5.1.13), (5.1.19b) and (5.1.36) yield
$$
\Xi=\Lambda .
\eqno (5.1.42)
$$
The anti-self-dual solution of Einstein's equations is thus
given by (5.1.30), (5.1.35) and (5.1.42).

The construction of an anti-self-dual space-time described in
this section is a particular example of the so-called non-linear
graviton \cite{Penr76a}, \cite{Penr76b}. 
In mathematical language, if $\cal M$
is a complex three-manifold, $B$ is the bundle of holomorphic 
three-forms on $\cal M$ and $H$ is the standard positive line
bundle on $P_{1}$, a non-linear graviton is the following set
of data \cite{Hitc79}:

(i) $\cal M$, the total space of a holomorphic fibration
$\pi: {\cal M} \rightarrow P_{1}$;

(ii) a four-parameter family of sections, each having 
$H \oplus H$ as normal bundle (see e.g. \cite{HuTo85} 
for the definition of normal bundle);

(iii) a non-vanishing holomorphic section $s$ of
$B \otimes \pi^{*} H^{4}$, where 
$H^{4} \equiv H\otimes H \otimes H
\otimes H$, and $\pi^{*}H^{4}$ denotes the pull-back of 
$H^{4}$ by $\pi$;

(iv) a real structure on $\cal M$ such that $\pi$ and $s$ are real.
$\cal M$ is then fibred from the real sections of the family.

\subsection{Beyond anti-self-duality}

The limit of the analysis performed in Sec. 5.1 is that it
deals with a class of solutions of (complex) Einstein equations
which is not sufficiently general. In \cite{YaIs82}
and \cite{Yass87} the authors have examined in detail
the limits of the anti-self-dual analysis.
The two main criticisms are as follows:

(a) a right-flat space-time (cf. the analysis in \cite{Law85}) 
does not represent a real Lorentzian
space-time manifold. Hence it cannot be applied directly to
classical gravity \cite{Ward80b};

(b) there are reasons fo expecting that the equations of a
quantum theory of gravity are much more complicated, and thus
are not solved by right-flat space-times.
\vskip 0.3cm
\noindent
However, an alternative approach due to Le Brun has become
available in the eighties \cite{Lebr85}. Le Brun's approach
focuses on the space $G$ of complex null geodesics of
complex space-time $(M,g)$, 
called ambitwistor space. Thus, one deals
with a standard rank-2 holomorphic vector bundle 
$E \rightarrow G$, and in the conformal class determined by
the complex structure of $G$, a one-to-one correspondence
exists between non-vanishing holomorphic sections of $E$ and
Einstein metrics on $(M,g)$ \cite{Lebr85}. The bundle $E$
is called Einstein bundle, and has also been studied in 
\cite{East87}. The work by Eastwood adds evidence in favour
of the Einstein bundle being the correct generalization of
the non-linear-graviton construction to the non-right-flat
case (cf. \cite{Law85}, \cite{Park90}, \cite{Lebr91}, \cite{Park91}).
Indeed, the theorems discussed so far provide a 
characterization of the vacuum Einstein equations. However,
there is not yet an independent way of recognizing the
Einstein bundle. Thus, this is not yet a substantial progress
in solving the vacuum equations. 
Other relevant work on holomorphic ideas appears 
in \cite{Lebr86}, where the author proves that,
in the case of four-manifolds with self-dual Weyl curvature,
solutions of the Yang--Mills equations correspond to 
holomorphic bundles on an associated analytic space
(cf. \cite{Ward77}, \cite{Witt78}, \cite{Ward81a}).

\subsection{Twistors as spin-${3\over 2}$ charges}

In this section, we describe a proposal by Penrose
to regard twistors for Ricci-flat space-times as (conserved)
{\it charges} for massless helicity-${3\over 2}$ fields
(\cite{Penr90}, \cite{Penr91a}, \cite{Penr91b}, \cite{Penr91c}).
The new approach proposed
by Penrose is based on the following mathematical results
\cite{Penr91b}:

(i) A vanishing Ricci tensor provides the consistency condition
for the existence and propagation of massless 
helicity-${3\over 2}$ fields in curved space-time
(\cite{Buch58}, \cite{DeZu76});

(ii) In Minkowski space-time, the space of charges for such 
fields is naturally identified with the corresponding twistor
space.
\vskip 0.3cm
\noindent
Thus, Penrose points out that if one could find the appropriate
definition of charge for massless helicity-${3\over 2}$ fields
in a Ricci-flat space-time, this should provide the concept of
twistor appropriate for vacuum Einstein equations. The corresponding
geometric program may be summarized as follows:

(1) Define a twistor for Ricci-flat space-time
$(M,g)_{RF}$;

(2) Characterize the resulting twistor space $\cal F$;

(3) Reconstruct $(M,g)_{RF}$ from $\cal F$.
\vskip 0.3cm
\noindent
We now describe, following \cite{Penr90}, 
\cite{Penr91a}, \cite{Penr91b}, \cite{Penr91c},
properties and problems of this approach to twistor theory
in flat and in curved space-times.
\vskip 1cm
\centerline {\bf 5.3.1 Massless spin-${3\over 2}$ equations
in Minkowski space-time}
\vskip 1cm
\noindent
Let $(M,\eta)$ be Minkowski space-time with flat connection
$\cal D$. In $(M,\eta)$ the gauge-invariant field strength
for spin ${3\over 2}$ is represented by a totally symmetric
spinor field 
$$
\psi_{A'B'C'}=\psi_{(A'B'C')},
\eqno (5.3.1)
$$
obeying a massless free-field equation
$$
{\cal D}^{AA'} \; \psi_{A'B'C'}=0.
\eqno (5.3.2)
$$
With the conventions of Penrose, $\psi_{A'B'C'}$ describes
spin-${3\over 2}$ particles of helicity equal to
${3\over 2}$ (rather than -${3\over 2}$). The {\it Dirac form}
of this field strength is obtained by expressing {\it locally}
$\psi_{A'B'C'}$ in terms of two potentials subject to gauge
freedoms involving a primed and an unprimed spinor field.
The first potential is a spinor field symmetric in its
primed indices
$$
\gamma_{B'C'}^{A}=\gamma_{(B'C')}^{A} ,
\eqno (5.3.3)
$$
subject to the differential equation
$$
{\cal D}^{BB'} \; \gamma_{B'C'}^{A}=0,
\eqno (5.3.4)
$$
and such that
$$
\psi_{A'B'C'}={\cal D}_{AA'} \; \gamma_{B'C'}^{A}.
\eqno (5.3.5)
$$
The second potential is a spinor field symmetric in its
unprimed indices
$$
\rho_{C'}^{AB}=\rho_{C'}^{(AB)},
\eqno (5.3.6)
$$
subject to the equation
$$
{\cal D}^{CC'} \; \rho_{C'}^{AB}=0,
\eqno (5.3.7)
$$
and it yields the $\gamma_{B'C'}^{A}$ potential by
means of
$$
\gamma_{B'C'}^{A}={\cal D}_{BB'} \; \rho_{C'}^{AB}.
\eqno (5.3.8)
$$
If we introduce the spinor fields $\nu_{C'}$ and $\chi^{B}$
obeying the equations
$$
{\cal D}^{AC'} \; \nu_{C'}=0,
\eqno (5.3.9)
$$
$$
{\cal D}_{AC'} \; \chi^{A}=2i \; \nu_{C'},
\eqno (5.3.10)
$$
the gauge freedoms for the two potentials enable one
to replace them by the potentials
$$
{\widehat \gamma}_{B'C'}^{A} \equiv \gamma_{B'C'}^{A}+
{\cal D}_{B'}^{\; \; \; A} \; \nu_{C'},
\eqno (5.3.11)
$$
$$
{\widehat \rho}_{C'}^{AB} \equiv \rho_{C'}^{AB}
+\varepsilon^{AB} \; \nu_{C'}
+i \; {\cal D}_{C'}^{\; \; \; A} \; \chi^{B},
\eqno (5.3.12)
$$
without affecting the theory. Note that the right-hand side of
(5.3.12) does not contain antisymmetric parts since, despite
the explicit occurrence of the antisymmetric $\varepsilon^{AB}$,
one finds
$$
{\cal D}_{C'}^{\; \; \; [A} \; \chi^{B]}
={\varepsilon^{AB} \over 2}{\cal D}_{LC'} \; \chi^{L}
=i \; \varepsilon^{AB} \nu_{C'},
\eqno (5.3.13)
$$
by virtue of (5.3.10). Hence (5.3.13) leads to
$$
{\widehat \rho}_{C'}^{AB}=\rho_{C'}^{AB}
+i \; {\cal D}_{C'}^{\; \; \; (A} \; \chi^{B)} .
\eqno (5.3.14)
$$
The gauge freedoms are indeed given by Eqs. (5.3.11) and (5.3.12)
since in our flat space-time one finds
$$
{\cal D}^{AA'} \; {\widehat \gamma}_{A'B'}^{C}=
{\cal D}^{AA'} \; {\cal D}_{\; \; B'}^{C} \; \nu_{A'}=
{\cal D}_{\; \; B'}^{C} \; {\cal D}^{AA'} \; \nu_{A'}=0 ,
\eqno (5.3.15)
$$
by virtue of (5.3.4) and (5.3.9), and
$$ 
{\cal D}^{AA'} \; {\widehat \rho}_{A'}^{BC}=
{\cal D}^{AA'} \; {\cal D}_{\; \; A'}^{C} \; \chi^{B}
={\cal D}^{CA'} \; {\cal D}_{A'}^{\; \; \; A} \; \chi^{B}
$$
$$
={\cal D}_{A'}^{\; \; \; A} \; {\cal D}^{CA'} \; \chi^{B}
=-{\cal D}^{AA'} \; {\cal D}_{\; \; A'}^{C} \; \chi^{B} ,
\eqno (5.3.16a) 
$$
which implies
$$
{\cal D}^{AA'} \; {\widehat \rho}_{A'}^{BC}=0 .
\eqno (5.3.16b)
$$
The result (5.3.16b) is a particular case of the application
of spinor Ricci identities to flat space-time.

We are now in a position to show that twistors can be regarded
as charges for helicity-${3\over 2}$ massless fields in Minkowski
space-time. For this purpose, following \cite{Penr91a},  
\cite{Penr91c} let us
suppose that the field $\psi$ satisfying (5.3.1) and (5.3.2) exists
in a region ${\cal R}$ of $(M,\eta)$, surrounding a world-tube
which contains the sources for $\psi$. Moreover, we consider a
two-sphere $\cal S$ within $\cal R$ surrounding the world-tube. 
To achieve this we begin by taking a {\it dual} twistor, i.e. the
pair of spinor fields
$$
W_{\alpha} \equiv \Bigr(\lambda_{A},\mu^{A'}\Bigr) ,
\eqno (5.3.17)
$$
obeying the differential equations
$$
{\cal D}_{AA'} \; \mu^{B'}=i \; \varepsilon_{A'}^{\; \; \; B'}
\; \lambda_{A} ,
\eqno (5.3.18)
$$
$$
{\cal D}_{AA'} \; \lambda_{B}=0.
\eqno (5.3.19)
$$
Hence $\mu^{B'}$ is a solution of the complex-conjugate twistor
equation
$$
{\cal D}_{A}^{(A'} \; \mu^{B')}=0.
\eqno (5.3.20)
$$
Thus, if one defines
$$
\varphi_{A'B'} \equiv \psi_{A'B'C'} \; \mu^{C'},
\eqno (5.3.21)
$$
one finds, by virtue of (5.3.1), (5.3.2) and (5.3.20), that
$\varphi_{A'B'}$ is a solution of the self-dual vacuum
Maxwell equations
$$
{\cal D}^{AA'} \; \varphi_{A'B'}=0.
\eqno (5.3.22)
$$
Note that (5.3.21) is a particular case of the spin-lowering
procedure \cite{HuTo85}, \cite{PeRi86}.  
Moreover, $\varphi_{A'B'}$ enables one to define the self-dual
two-form
$$
F \equiv \varphi_{A'B'} \; dx_{A}^{\; \; A'} \; \wedge \;
dx^{AB'},
\eqno (5.3.23)
$$
which leads to the following {\it charge} assigned to the
world-tube:
$$
Q \equiv {i\over 4\pi} \oint F .
\eqno (5.3.24)
$$
For some twistor
$$
Z^{\alpha} \equiv \Bigr(\omega^{A},\pi_{A'}\Bigr),
\eqno (5.3.25)
$$
the charge $Q$ depends on the dual twistor $W_{\alpha}$ as
$$
Q =Z^{\alpha} \; W_{\alpha}=\omega^{A} \; \lambda_{A}
+\pi_{A'} \; \mu^{A'}.
\eqno (5.3.26)
$$
These equations describe the strength of the charge, for the
field $\psi$, that should be assigned to the world-tube.
Thus, a twistor $Z^{\alpha}$ arises naturally in Minkowski
space-time as the charge for a helicity $+{3\over 2}$ massless
field, whereas a dual twistor $W_{\alpha}$ is the charge for a
helicity $-{3\over 2}$ massless field \cite{Penr91c}.

Interestingly, the potentials $\gamma_{A'B'}^{C}$ and 
$\rho_{A'}^{BC}$ can be used to obtain a potential for the
self-dual Maxwell field strength, since, after defining
$$
\theta_{\; \; A'}^{C} \equiv \gamma_{A'B'}^{C} \; \mu^{B'}
-i \; \rho_{A'}^{BC} \; \lambda_{B},
\eqno (5.3.27)
$$
one finds
$$ 
{\cal D}_{CB'} \; \theta_{\; \; A'}^{C} =
\Bigr({\cal D}_{CB'} \; \gamma_{A'D'}^{C}\Bigr)\mu^{D'}
+\gamma_{A'D'}^{C} \Bigr({\cal D}_{CB'} \; \mu^{D'}\Bigr)
-i \Bigr({\cal D}_{CB'} \; \rho_{A'}^{BC}\Bigr)\lambda_{B} 
$$
$$
=\psi_{A'B'D'} \; \mu^{D'}+i \; \varepsilon_{B'}^{\; \; \; D'} \;
\gamma_{A'D'}^{C} \; \lambda_{C}
-i \; \gamma_{A'B'}^{C} \; \lambda_{C}
$$
$$
=\psi_{A'B'D'} \; \mu^{D'}=\varphi_{A'B'},
\eqno (5.3.28) 
$$
$$ 
{\cal D}_{B}^{\; \; A'} \; \theta_{\; \; A'}^{C}=
\Bigr({\cal D}_{B}^{\; \; A'} \; \gamma_{A'B'}^{C}\Bigr)
\mu^{B'}+\gamma_{A'B'}^{C}\Bigr({\cal D}_{B}^{\; \; A'}
\; \mu^{B'}\Bigr)-i\Bigr({\cal D}_{B}^{\; \; A'} \;
\rho_{A'}^{DC}\Bigr) \lambda_{D}
$$
$$
-i \rho_{A'}^{DC} \Bigr({\cal D}_{B}^{\; \; A'} 
\; \lambda_{D}\Bigr)=0 .
\eqno (5.3.29) 
$$
Eq. (5.3.28) has been obtained by using (5.3.5), (5.3.8),
(5.3.18) and (5.3.19), whereas (5.3.29) holds by virtue of
(5.3.3), (5.3.4), (5.3.7), (5.3.18) and (5.3.19). The one-form
corresponding to $\theta_{\; \; A'}^{C}$ is defined by
$$
A \equiv \theta_{BB'} \; dx^{BB'} ,
\eqno (5.3.30)
$$
which leads to
$$
F=2 \; dA ,
\eqno (5.3.31)
$$
by using (5.3.23) and (5.3.28).

The {\it Rarita--Schwinger form} of the field strength does not
require the symmetry (5.3.3) in $B'C'$ as we have done so far,
and the $\gamma_{B'C'}^{A}$ potential is instead subject to the
equations \cite{Penr91a}, \cite{Penr91b}, \cite{Penr91b} 
$$
\varepsilon^{B'C'} \; {\cal D}_{A(A'} \; \gamma_{B')C'}^{A}
=0 ,
\eqno (5.3.32)
$$
$$
{\cal D}^{B'(B} \; \gamma_{B'C'}^{A)}=0 .
\eqno (5.3.33)
$$
Moreover, the spinor field $\nu_{C'}$ in (5.3.11) is no
longer taken to be a solution of the Weyl equation (5.3.9).

The potentials $\gamma$ and $\rho$ may or may not be global 
over $\cal S$. If $\gamma$ is global but $\rho$ is not, one
obtains a two-dimensional complex vector space parametrized 
by the spinor field $\pi_{A'}$. The corresponding subspace
where $\pi_{A'}=0$, parametrized by $\omega^{A}$, is called
$\omega$-space. Thus, following \cite{Penr91c}, we regard
$\pi$-space and $\omega$-space as quotient spaces defined 
as follows:
$$
\pi-{\rm space} \equiv {\rm space} 
\; {\rm of} \; {\rm global} \; 
\psi{\rm 's} / {\rm space} \; {\rm of} \; {\rm global}
\; \gamma{\rm 's},
\eqno (5.3.34)
$$
$$
\omega-{\rm space} \equiv {\rm space} 
\; {\rm of} \; {\rm global} \;
\gamma{\rm 's} / {\rm space} \; {\rm of} \; 
{\rm global} \; \rho{\rm 's}.
\eqno (5.3.35)
$$
\vskip 1cm
\centerline {\bf 5.3.2 Massless spin-${3\over 2}$ field strengths
in curved space-time}
\vskip 1cm
\noindent
The conditions for the {\it local} existence of the
$\rho_{A'}^{BC}$ potential in curved space-time are derived
by requiring that, after the gauge transformation (5.3.12)
(or, equivalently, (5.3.14)), also the 
${\widehat \rho}_{A'}^{BC}$ potential should obey the equation
$$
\nabla^{AA'} \; {\widehat \rho}_{A'}^{BC}=0,
\eqno (5.3.36)
$$
where $\nabla$ is the curved connection. 
By virtue of the spinor Ricci identity \cite{WaWe90}
$$
\nabla_{M'(A} \; \nabla_{\; \; \; B)}^{M'} \; \chi_{C}
=\psi_{ABDC} \; \chi^{D}-2\Lambda \; \chi_{(A} \;
\varepsilon_{B)C},
\eqno (5.3.37)
$$
the insertion of (5.3.14) into (5.3.36) yields, assuming
for simplicity that $\nu_{C'}=0$ in (5.3.10), the following
conditions:
$$
\psi_{ABCD}=0, \;
\Lambda=0 ,
\eqno (5.3.38)
$$
which imply we deal with a vacuum self-dual (or left-flat)
space-time, since the anti-self-dual Weyl spinor has
to vanish \cite{Penr91c}.

Moreover, in a complex anti-self-dual vacuum space-time 
one finds \cite{Penr91c} that spin-${3\over 2}$ field
strengths $\psi_{A'B'C'}$ can be defined according to
(cf. (5.3.5))
$$
\psi_{A'B'C'}=\nabla_{CC'} \; \gamma_{A'B'}^{C},
\eqno (5.3.39)
$$
are gauge-invariant, totally symmetric, and satisfy the massless
free-field equations (cf. (5.3.2))
$$
\nabla^{AA'} \; \psi_{A'B'C'}=0.
\eqno (5.3.40)
$$
In this case there is no obstruction to defining global 
$\psi$-fields with non-vanishing $\pi$-charge, and a global
$\pi$-space can be defined as in (5.3.34). It remains to be
seen whether the twistor space defined by $\alpha$-surfaces
may then be reconstructed (Sec. 4.2, \cite{Penr76a}, \cite{Penr76b},
\cite{WaWe90}, \cite{Penr91c}).

Interestingly, in \cite{Penr91b} it has been proposed to
interpret the potential $\gamma$ as providing a 
{\it bundle connection}. In other words, one takes the
fibre coordinates to be given by a spinor $\eta_{A'}$ and
a scalar $\mu$. For a given small $\epsilon$, one extends
the ordinary Levi--Civita connection $\nabla$ on $M$ to
bundle-valued quantities according to \cite{Penr91b}
$$
\nabla_{PP'} \pmatrix {\eta_{A'} \cr \mu \cr}
\equiv
\pmatrix {\nabla_{PP'} \; \eta_{A'} \cr
\nabla_{PP'} \; \mu}
-\epsilon \pmatrix {0 & \gamma_{PP'A'} \cr
\gamma_{PP'}^{\; \; \; \; \; \; B'} & 0 \cr}
\pmatrix {\eta_{B'} \cr \mu \cr},
\eqno (5.3.41)
$$
with gauge transformations given by
$$
\pmatrix {{\widehat \eta}_{A'} \cr {\widehat \mu} \cr}
\equiv
\pmatrix {\eta_{A'} \cr \mu \cr}
+\epsilon \pmatrix {0 & \nu_{A'} \cr
\nu^{B'} & 0 \cr}
\pmatrix {\eta_{B'} \cr \mu \cr}.
\eqno (5.3.42)
$$
Note that terms of order $\epsilon^{2}$ have been
neglected in writing (5.3.42). However, such gauge
transformations do not close under commutation, and to
obtain a theory valid to all orders in $\epsilon$ one has
to generalize to $SL(3,C)$ matrices before the commutators
close. Writing $(A)$ for the three-dimensional 
indices, so that $\eta_{(A)}$ denotes
$\pmatrix {\eta_{A'} \cr \mu \cr}$, one has a connection
defined by
$$
\nabla_{PP'} \; \eta_{(A)} \equiv
\pmatrix {\nabla_{PP'} \; \eta_{A'} \cr
\nabla_{PP'} \; \mu \cr}
-\gamma_{PP' \; (A)} 
^{\; \; \; \; \; \; \; \; \; \; \; (B)}
\; \; \eta_{(B)},
\eqno (5.3.43)
$$
with gauge transformation
$$
{\widehat \eta}_{(A)} \equiv
\eta_{(A)}
+\nu_{(A)}^{\; \; \; \; (B)}
\; \eta_{(B)} .
\eqno (5.3.44)
$$
With this notation, the 
$\nu_{(A)}^{\; \; \; \; (B)}$ are 
$SL(3,C)$-valued fields on $M$, and hence
$$
{\cal E}^{(P) \; (Q) \; (R)} \; \;
\nu_{(P)}^{\; \; \; \; (A)}
\; \; \nu_{(Q)}^{\; \; \; \; (B)}
\; \; \nu_{(R)}^{\; \; \; \; (C)}
= {\cal E}^{(A) \; (B) \; (C)},
\eqno (5.3.45)
$$
where ${\cal E}^{(P) \; (Q) \; (R)}$
are generalized Levi--Civita 
symbols. The $SL(3,C)$ definition of $\gamma$-potentials
takes the form \cite{Penr91b}
$$
\gamma_{PP' \; (A)}^{\; \; \; \; \; \; \; \; \; \; \; (B)}
\equiv
\pmatrix {\alpha_{PP'A'}^{\; \; \; \; \; \; \; \; \; \; B'}
& \beta_{PP'A'} \cr
\gamma_{PP'}^{\; \; \; \; \; \; B'} & \delta_{PP'} \cr},
\eqno (5.3.46)
$$
while the curvature is
$$
K_{pq \; (A)}^{\; \; \; \; \; \; \; \; (B)}
\equiv 2 \nabla_{[p} \; 
\gamma_{q](A)}^{\; \; \; \; \; \; \; (B)}
+2 \; \gamma_{[p \mid (A) \mid }
^{\; \; \; \; \; \; \; \; \; (C)}
\; \; \gamma_{q](C)}^{\; \; \; \; \; \; \; \; (B)}.
\eqno (5.3.47)
$$
Penrose has proposed this as a generalization of the
Rarita--Schwinger structure in Ricci-flat space-times, and he
has even speculated that a non-linear generalization of the
Rarita--Schwinger equations (5.3.32) and (5.3.33) might be
$$
{ }^{(-)}K_{PQ \; (A)}^{\; \; \; \; \; \; \; \; \; \; (B)}=0,
\eqno (5.3.48)
$$
$$
{ }^{(+)}K_{P'Q' \; (A)} 
^{\; \; \; \; \; \; \; \; \; \; \; \; (B)}
\; \; {\cal E}^{P' \; (A) \; (C)}
\; \;  
{\cal E}_{\; \; \; \; (B) \; (D)}^{Q'}=0,
\eqno (5.3.49)
$$
where ${ }^{(-)}K$ and ${ }^{(+)}K$ are the anti-self-dual
and self-dual parts of the curvature respectively, i.e.
$$
K_{pq \; (A)}^{\; \; \; \; \; \; \; \; \; (B)}
= \varepsilon_{P'Q'} \; \; 
{ }^{(-)}K_{PQ \; (A)}^{\; \; \; \; \; 
\; \; \; \; \; (B)}
+ \varepsilon_{PQ} \; \;
{ }^{(+)}K_{P'Q' \; (A)}^{\; \; \; \; \; 
\; \; \; \; \; \; \; \; \; (B)}.
\eqno (5.3.50)
$$
Following \cite{Penr91b}, one has
$$
{\cal E}^{P' \; (A) \; (C)}
\equiv {\cal E}^{(P) \; (A)
\; (C)}
\; e_{(P)}^{\; \; \; \; P'},
\eqno (5.3.51)
$$
$$
{\cal E}_{Q' \; (B) \; (D)}
\equiv {\cal E}_{(Q) \; (B) \; (D)}
\; \; e_{Q'}^{\; \; \; \; (Q)},
\eqno (5.3.52)
$$
the $e_{(P)}^{\; \; \; \; P'}$ and
$e_{Q'}^{\; \; \; (Q)}$ relating the bundle directions
with tangent directions in $M$. 

\section{The Plebanski Contributions}

The analysis of (conformally) right-flat 
space-times of the previous sections has its counterpart in the
theory of heaven spaces developed by Plebanski. This section
reviews weak heaven spaces, strong heaven spaces,
heavenly tetrads and heavenly equations. 

\subsection{Outline}

One of the most recurring themes of this paper is the
analysis of complex or real Riemannian manifolds where half
of the conformal curvature vanishes and the vacuum Einstein
equations hold. Section 5 has provided an explicit construction
of such anti-self-dual space-times, and the underlying 
Penrose-transform theory has been presented in Sec. 4.
However, alternative ways exist to construct these
solutions of the Einstein equations, and hence this section
supplements the previous sections by describing the work
in \cite{Pleb75}. By using the tetrad formalism and some
basic results in the theory of partial differential equations,
the so-called {\it heaven spaces} and {\it heavenly tetrads}
are defined and constructed in detail. 

\subsection{Heaven spaces}

In his theory of heaven spaces, Plebanski studies a 
four-dimensional {\it analytic} manifold $M_{4}$ with metric
given in terms of tetrad vectors as \cite{Pleb75}
$$
g=2e^{1}e^{2}+2e^{3}e^{4}=g_{ab} \; e^{a} e^{b}
\; \in \Lambda^{1} \otimes \Lambda^{1}.
\eqno (6.2.1)
$$
The definition of the $2 \times 2$ matrices
$$
\tau^{AB'} \equiv \sqrt{2} 
\pmatrix {e^{4}&e^{2}\cr e^{1}&-e^{3} \cr}
\eqno (6.2.2)
$$
enables one to re-express the metric as
$$
g=-{\rm det} \; \tau^{AB'}
={1\over 2} \; \varepsilon_{AB} \; \varepsilon_{C'D'} \;
\tau^{AC'} \; \tau^{BD'}.
\eqno (6.2.3)
$$
Moreover, since the manifold is analytic, there exist two
{\it independent} sets of $2 \times 2$ complex matrices
with unit determinant: $L_{\; \; \; A}^{A'} \in \; SL(2,C)$
and ${\widetilde L}_{\; \; \; B}^{B'} \in \;
{\widetilde {SL}}(2,C)$. On defining a new set of tetrad vectors
such that
$$
\sqrt{2} \pmatrix {e^{4'}&e^{2'}\cr e^{1'}&-e^{3'}\cr}
=L_{\; \; \; A}^{A'} \;
{\widetilde L}_{\; \; \; B'}^{B}
\; \tau^{AB'},
\eqno (6.2.4)
$$
the metric is still obtained as $2e^{1'}e^{2'}+2e^{3'}e^{4'}$.
Hence the tetrad gauge group may be viewed as
$$
{\cal G} \equiv SL(2,C) \times {\widetilde {SL}}(2,C).
\eqno (6.2.5)
$$

A key role in the following analysis is played by a pair of
differential forms whose spinorial version is obtained from
the wedge product of the matrices in (6.2.2), i.e.
$$
\tau^{AB'}\wedge \tau^{CD'}=S^{AC} \; \varepsilon^{B'D'}
+\varepsilon^{AC} \; {\widetilde S}^{B'D'},
\eqno (6.2.6)
$$
where
$$
S^{AB} \equiv {1\over 2} \; \varepsilon_{R'S'} \;
\tau^{AR'} \wedge \tau^{BS'}
={1\over 2} \; e^{a}\wedge e^{b} \; S_{ab}^{\; \; \; AB},
\eqno (6.2.7)
$$
$$
{\widetilde S}^{A'B'} \equiv {1\over 2} \;
\varepsilon_{RS} \; \tau^{RA'} \wedge \tau^{SB'}
={1\over 2} \; e^{a} \wedge e^{b} \;
{\widetilde S}_{ab}^{\; \; \; A'B'}.
\eqno (6.2.8)
$$
The forms $S^{AB}$ and ${\widetilde S}^{A'B'}$ are self-dual
and anti-self-dual respectively, in that the action of the 
Hodge-star operator on them leads to \cite{Pleb75}
$$
{ }^{*}S^{AB}=S^{AB},
\eqno (6.2.9)
$$
$$
{ }^{*}{\widetilde S}^{A'B'}=-{\widetilde S}^{A'B'}.
\eqno (6.2.10)
$$
To obtain the desired spinor description of the curvature, we
introduce the antisymmetric connection forms
$\Gamma_{ab}=\Gamma_{[ab]}$ through the first structure
equations
$$
de^{a}=e^{b} \wedge \Gamma_{\; \; b}^{a}.
\eqno (6.2.11)
$$
The spinorial counterpart of $\Gamma_{ab}$ is given by
$$
\Gamma_{AB} \equiv -{1\over 4} \Gamma_{ab} \; 
S_{\; \; \; AB}^{ab},
\eqno (6.2.12)
$$
$$
{\widetilde \Gamma}_{A'B'} \equiv -{1\over 4}
\Gamma_{ab} \; {\widetilde S}_{\; \; \; A'B'}^{ab},
\eqno (6.2.13)
$$
which implies
$$
\Gamma_{ab}=-{1\over 2}S_{ab}^{\; \; \; AB} 
\; \Gamma_{AB} -{1\over 2} {\widetilde S}_{ab}^{\; \; \; A'B'}
\; {\widetilde \Gamma}_{A'B'}.
\eqno (6.2.14)
$$
To appreciate that $\Gamma_{AB}$ and ${\widetilde \Gamma}_{A'B'}$
are actually independent, the reader may find it useful to check
that \cite{Pleb75}
$$
\Gamma_{AB}=-{1\over 2} \pmatrix 
{2\Gamma_{42}& {\Gamma_{12}+\Gamma_{34}}\cr
{\Gamma_{12}+\Gamma_{34}} & 2\Gamma_{31}\cr},
\eqno (6.2.15)
$$
$$
{\widetilde \Gamma}_{A'B'}=-{1\over 2} \pmatrix
{2\Gamma_{41} & {-\Gamma_{12}+\Gamma_{34}}\cr
{-\Gamma_{12}+\Gamma_{34}}& 2\Gamma_{32}\cr}.
\eqno (6.2.16)
$$
The action of exterior differentiation on
$\tau^{AB'},S^{AB},{\widetilde S}^{A'B'}$ shows that
$$
d\tau^{AB'}=\tau^{AL'} 
\wedge {\widetilde \Gamma}_{\; \; \; L'}^{B'}
+\tau^{LB'} \wedge \Gamma_{\; \; L}^{A},
\eqno (6.2.17)
$$
$$
dS^{AB}=-3S^{(AB} \; \Gamma_{\; \; \; C}^{C)},
\eqno (6.2.18)
$$
$$
d{\widetilde S}^{A'B'}=-3{\widetilde S}^{(A'B'} \; 
{\widetilde \Gamma}_{\; \; \; \; C'}^{C')},
\eqno (6.2.19)
$$
and two {\it independent} curvature forms are obtained as
$$ 
R_{\; \; B}^{A} \equiv
d\Gamma_{\; \; B}^{A}+\Gamma_{\; \; L}^{A} \wedge 
\Gamma_{\; \; B}^{L} 
=-{1\over 2} \; \psi_{\; \; BCD}^{A} \; S^{CD}
$$
$$
\; \; +{R\over 24} \; S_{\; \; B}^{A}
+{1\over 2} \; \Phi_{\; \; BC'D'}^{A}
\; {\widetilde S}^{C'D'},
\eqno (6.2.20) 
$$
$$ 
{\widetilde R}_{\; \; \; B'}^{A'}  \equiv
d{\widetilde \Gamma}_{\; \; \; B'}^{A'}
+{\widetilde \Gamma}_{\; \; \; L'}^{A'}
\wedge {\widetilde \Gamma}_{\; \; \; B'}^{L'}
=-{1\over 2} \; {\widetilde \psi}_{\; \; \; B'C'D'}^{A'}
\; {\widetilde S}^{C'D'}
$$
$$
\; \; +{R\over 24} \; 
{\widetilde S}_{\; \; \; B'}^{A'}
+{1\over 2} \; \Phi_{CD \; \; \; \; B'}^{\; \; \; \; \; \; A'}
\; S^{CD} .
\eqno (6.2.21)
$$
The spinors and scalars in (6.2.20) and (6.2.21) have the same
meaning as in the previous sections. With the conventions in
\cite{Pleb75}, the Weyl spinors are obtained as
$$
\psi_{ABCD}={1\over 16} \; S_{\; \; \; AB}^{ab}
\; C_{abcd} \; S_{\; \; \; CD}^{cd}
=\psi_{(ABCD)} ,
\eqno (6.2.22)
$$
$$
{\widetilde \psi}_{A'B'C'D'}={1\over 16} \;
{\widetilde S}_{\; \; \; A'B'}^{ab}
\; C_{abcd} \;
{\widetilde S}_{\; \; \; C'D'}^{cd}
={\widetilde \psi}_{(A'B'C'D')} ,
\eqno (6.2.23)
$$
and conversely the Weyl tensor is
$$
C_{abcd}={1\over 4} \; S_{ab}^{\; \; \; AB}
\; \psi_{ABCD} \; S_{cd}^{\; \; \; CD}
+{1\over 4} \; {\widetilde S}_{ab}^{\; \; \; A'B'}
\; {\widetilde \psi}_{A'B'C'D'}
\; {\widetilde S}_{cd}^{\; \; \; C'D'}.
\eqno (6.2.24)
$$
The spinor version of the Petrov classification 
(Sec. 2.3) is hence obtained
by stating that $k^{A}$ and $\omega^{A'}$ are the two types of
P-spinors if and only if the {\it independent} conditions hold:
$$
\psi_{ABCD} \; k^{A} \; k^{B} \; k^{C} \; k^{D}=0,
\eqno (6.2.25)
$$
$$
{\widetilde \psi}_{A'B'C'D'} \; \omega^{A'} \;
\omega^{B'} \; \omega^{C'} \; \omega^{D'}=0.
\eqno (6.2.26)
$$
For our purposes, we can omit the details about the principal null
directions, and focus instead on the classification of spinor fields
and analytic manifolds under consideration. Indeed, Plebanski
proposed to call all objects which are ${\widetilde {SL}}(2,C)$
scalars and are geometric objects with respect to $SL(2,C)$,
the {\it heavenly objects} (e.g. $S^{AB},\Gamma_{AB},\psi_{ABCD}$).
Similarly, objects which are $SL(2,C)$ scalars and behave like
geometric objects with respect to ${\widetilde {SL}}(2,C)$ belong
to the complementary world, i.e. the set of {\it hellish objects}
(e.g. ${\widetilde S}^{A'B'},{\widetilde \Gamma}_{A'B'},
{\widetilde \psi}_{A'B'C'D'}$). Last, spinor fields with (abstract)
indices belonging to both primed and unprimed spin-spaces are the
{\it earthly objects}.

With the terminology of Plebanski, a {\it weak heaven} space is
defined by the condition 
$$
{\widetilde \psi}_{A'B'C'D'}=0,
\eqno (6.2.27)
$$
and corresponds to the {\it conformally right-flat} space
of Sec. 3. Moreover, a {\it strong heaven} space is
a four-dimensional analytic manifold where a choice of null
tetrad exists such that
$$
{\widetilde \Gamma}_{A'B'}=0.
\eqno (6.2.28)
$$
One then has {\it a forteriori}, by virtue of (6.2.21),
the conditions \cite{Pleb75}
$$
{\widetilde \psi}_{A'B'C'D'}=0, \;
\Phi_{ABC'D'}=0, \;
R=0 .
\eqno (6.2.29)
$$
The vacuum Einstein equations are then automatically fulfilled 
in a strong heaven space, which turns out to be a right-flat
space-time in modern language. Of course, strong heaven spaces
are non-trivial if and only if the anti-self-dual Weyl spinor
$\psi_{ABCD}$ does not vanish, otherwise they reduce to flat
four-dimensional space-time.

\subsection{First heavenly equation}

A space which is a strong heaven according to (6.2.28) is
characterized by a key function $\Omega$ which obeys the
so-called first heavenly equation. The basic ideas are as
follows. In the light of (6.2.19) and (6.2.28),
$d{\widetilde S}^{A'B'}$ vanishes, and hence, {\it in a simply
connected region}, an element $U^{A'B'}$ of the bundle
$\Lambda^{1}$ exists such that locally
$$
{\widetilde S}^{A'B'}=dU^{A'B'}.
\eqno (6.3.1)
$$
Thus, since
$$
{\widetilde S}^{1'1'}=2 e^{4} \wedge e^{1},
\eqno (6.3.2)
$$
$$
{\widetilde S}^{2'2'}=2 e^{3} \wedge e^{2},
\eqno (6.3.3)
$$
$$
{\widetilde S}^{1'2'}=-e^{1} \wedge e^{2}
+e^{3} \wedge e^{4},
\eqno (6.3.4)
$$
Equation (6.3.1) leads to
$$
2 e^{4} \wedge e^{1}=dU^{1'1'},
\eqno (6.3.5)
$$
$$
2 e^{3} \wedge e^{2}=dU^{2'2'}.
\eqno (6.3.6)
$$
Now the Darboux theorem holds in our complex manifold, and
hence scalar functions $p,q,r,s$ exist such that
$$
2 e^{4} \wedge e^{1} = 2 dp \wedge dq = 2 d(p \; dq + d\tau),
\eqno (6.3.7)
$$
$$
2 e^{3} \wedge e^{2} = 2 dr \wedge ds = 2 d(r \; ds + d\sigma),
\eqno (6.3.8)
$$
$$
e^{1}\wedge e^{2} \wedge e^{3} \wedge e^{4}=
dp \wedge dq \wedge dr \wedge ds.
\eqno (6.3.9)
$$
The form of the {\it heavenly tetrad} in these coordinates is
$$
e^{1}=A \; dp +B \; dq,
\eqno (6.3.10)
$$
$$
e^{2}=G \; dr + H \; ds ,
\eqno (6.3.11)
$$
$$
e^{3}=E \; dr + F \; ds ,
\eqno (6.3.12)
$$
$$
e^{4}=-C \; dp -D \; dq .
\eqno (6.3.13)
$$
If one now inserts (6.3.10)--(6.3.13) into (6.3.7)--(6.3.9),
one finds that 
$$
AD-BC=EH-FG=1 ,
\eqno (6.3.14)
$$
which is supplemented by a set of equations resulting from the
condition $d{\widetilde S}^{1'2'}=0$. These
equations imply the existence of a function, the 
{\it first key function}, such that \cite{Pleb75}
$$
AG-CE=\Omega_{pr},
\eqno (6.3.15)
$$
$$
BG-DE=\Omega_{qr},
\eqno (6.3.16)
$$
$$
AH-CF=\Omega_{ps},
\eqno (6.3.17)
$$
$$
BH-DF=\Omega_{qs}.
\eqno (6.3.18)
$$
Thus, $E,F,G,H$ are given by
$$
E=B \; \Omega_{pr}-A \; \Omega_{qr},
\eqno (6.3.19)
$$
$$
F=B \; \Omega_{ps}-A \; \Omega_{qs},
\eqno (6.3.20)
$$
$$
G=D \; \Omega_{pr}-C \; \Omega_{qr},
\eqno (6.3.21)
$$
$$
H=D \; \Omega_{ps}-C \; \Omega_{qs}.
\eqno (6.3.22)
$$
The request of compatibility of (6.3.19)--(6.3.22) with (6.3.14)
leads to the {\it first heavenly equation}
$$
{\rm det} \; \pmatrix
{\Omega_{pr} & \Omega_{ps} \cr
\Omega_{qr} & \Omega_{qs} \cr}=1.
\eqno (6.3.23)
$$

\subsection{Second heavenly equation}

A more convenient description of the heavenly tetrad is obtained
by introducing the coordinates
$$
x \equiv \Omega_{p} , \;
y \equiv \Omega_{q},
\eqno (6.4.1)
$$
and then defining
$$
A \equiv -\Omega_{pp} , \;
B \equiv -\Omega_{pq}, \;
C \equiv -\Omega_{qq}.
\eqno (6.4.2)
$$
The corresponding heavenly tetrad reads \cite{Pleb75}
$$
e^{1}=dp ,
\eqno (6.4.3)
$$
$$
e^{2}=dx+A \; dp + B \; dq ,
\eqno (6.4.4)
$$
$$
e^{3}=-dy-B \; dp -C \; dq,
\eqno (6.4.5)
$$
$$
e^{4}=-dq.
\eqno (6.4.6)
$$
Now the closure condition for ${\widetilde S}^{2'2'}:
d{\widetilde S}^{2'2'}=0$, leads to the equations
$$
A_{x}+B_{y}=0,
\eqno (6.4.7)
$$
$$
B_{x}+C_{y}=0,
\eqno (6.4.8)
$$
$$
\Bigr(AC-B^{2}\Bigr)_{x}+B_{q}-C_{p}=0,
\eqno (6.4.9)
$$
$$
\Bigr(AC-B^{2}\Bigr)_{y}-A_{q}+B_{p}=0.
\eqno (6.4.10)
$$
By virtue of (6.4.7) and (6.4.8), a function $\theta$ exists
such that
$$
A=-\theta_{yy} , \;
B=\theta_{xy} , \;
C=-\theta_{xx}.
\eqno (6.4.11)
$$
On inserting (6.4.11) into (6.4.9) and (6.4.10) one finds
$$
\partial_{w} \Bigr(\theta_{xx} \; \theta_{yy}
-\theta_{xy}^{2}+\theta_{xp}+\theta_{yq}\Bigr)=0 ,
\eqno (6.4.12)
$$
where $w=x,y$. Thus, one can write that
$$
\theta_{xx}\theta_{yy}-\theta_{xy}^{2}+\theta_{xp}
+\theta_{yq}=f_{p}(p,q),
\eqno (6.4.13)
$$
where $f$ is an arbitrary function of $p$ and $q$. This 
suggests defining the function
$$
\Theta \equiv \theta-xf ,
\eqno (6.4.14)
$$
which implies
$$
f_{p}=\Theta_{xx}\Theta_{yy}-\Theta_{xy}^{2}
+\Theta_{xp}+\Theta_{yq}+f_{p},
$$
and hence
$$
\Theta_{xx} \; \Theta_{yy} -\Theta_{xy}^{2}
+\Theta_{xp}+\Theta_{yq}=0.
\eqno (6.4.15)
$$
Equation (6.4.15) ensures that all forms 
${\widetilde S}^{A'B'}$ are closed, and is called the
{\it second heavenly equation}. Plebanski was able to find
heavenly metrics of all possible algebraically degenerate
types. An example is given by the function
$$
\Theta \equiv {\beta \over 2\alpha (\alpha-1)}
x^{\alpha} \; y^{1-\alpha}.
\eqno (6.4.16)
$$
The reader may check that such a
solution is of the type $[2-2] \otimes [-]$ if 
$\alpha=-1,2$, and is of the type $[2-1-1] \otimes [-]$ 
whenever $\alpha \not = -1,2$ \cite{Pleb75}. More work on
related topics and on yet other ideas in complex general relativity
can be found in \cite{PlHa75}, \cite{FiPl76}, \cite{Newm76},
\cite{PlSc76}, \cite{Ko77},
\cite{Boye78}, \cite{Hans78}, \cite{Tod80},
\cite{ToWi80}, \cite{FiPl81}, \cite{Ko81},
\cite{SpTo81}, \cite{East90}, \cite{Lebr90}, 
\cite{BeSm91}, \cite{PlPr94},
\cite{PlGa95a}, \cite{PlGa95b}, \cite{ToDu97}.

\section{Concluding Remarks}

It has been our aim to give a pedagogical introduction to twistor
theory, with emphasis on the topics and methods that a general
relativist is more familiar with. Much more material can be found,
for example, in \cite{Espo95}, especially spin-3/2 potentials
(\cite{AiUr81}, \cite{Frau94}, \cite{MaPe94}, \cite{Frau95},
\cite{Frau96})
and various definitions of twistors in curved space-time
\cite{Penr75}. Moreover we should acknowledge that complex (Riemannian)
manifolds have been investigated from the point of view of the
corresponding real structure in \cite{Boro99}, \cite{Boro00}. These are 
the so-called Norden--K\"{a}hler or anti-K\"{a}hler manifolds. With spinor 
notation, they are treated in \cite{Robi02a}, \cite{Robi02b}. Older
references on this subject which deserve mention are \cite{GaIv92},
\cite{Ivan97}, \cite{Lebr82}, \cite{Flah76}, \cite{Flah78}, while very
recent results can be found in \cite{Mene04}, \cite{Sluk05}.

For a long time the physics community has thought that twistor theory
is more likely to contribute to mathematics, e.g. powerful geometric
methods for solving non-linear partial differential equations
(\cite{Ward81b}, \cite{Tod83}, \cite{Lewa91}, \cite{Penr94},
\cite{Tod95}, \cite{Nuro97}, \cite{Penr97}). However,
the work of Penrose on spin-3/2 potentials (\cite{Penr91c},
\cite{Penr94}) and the work in \cite{Witt04}
on perturbative gauge theory has changed a lot the overall perspectives.
In particular, Witten points out that perturbative scattering amplitudes
in Yang--Mills theories have remarkable properties such as holomorphy
of the maximally helicity violating amplitudes. To interpret this
result, he considers the Fourier transform of scattering amplitudes
from momentum space to twistor space, and finds that the transformed
amplitudes are supported on certain holomorphic curves. Hence he
suggests that this might result from an equivalence between the
perturbative expansion of N=4 super Yang--Mills theory and the
D-instanton expansion of the topological B-model the target space
of which is the Calabi--Yau supermanifold ${\bf CP}^{3/4}$. The subject
of twistor-string theory (see Twistor String Theory URL \cite{WWW05}
in the References) has evolved out of such a seminal paper,
showing once more the profound importance of holomorphic ideas
in quantum gravity.

Moreover, from the point of view of classical general relativity,
it appears very encouraging that general asymptotically flat (neither
necessarily self-dual nor anti-self-dual) vacuum four-spaces can be
described within a new twistor-geometric formalism \cite{Penr99}.

\section*{Acknowledgments}

Springer Science and Business Media have 
kindly granted permission to re-print 
the first five sections from the 1995 Kluwer book 
{\it Complex General Relativity} by the author.
The work of G. Esposito has been partially supported by PRIN 
{\it SINTESI}. He is indebted to G. Sardanashvily for
encouragement, and to Maciej Dunajski for enlightening conversations
and remarks.

\end{document}